\documentclass{elsart}

\usepackage{amsmath,amssymb}
\newcommand{\beginproof}{\noindent {\it Proof.}\ }

\newcommand{\eproof}{\hfill $\Box$}
\newcommand{\R}{{\mathbb R}}
\newcommand{\Z}{{\mathbb Z}}
\newcommand{\N}{{\mathbb N}}
\newcommand{\sP}{{\mathbb P}}

\newcommand{\bi}{\mathbf{i}}
\newcommand{\bj}{\mathbf{j}}

\newcommand{\mc}{\mathcal}
\newcommand{\diam}{\text{ \rm diam }}

\newcommand{\En}{{\mathcal E}_{s}}
\newcommand{\E}{E_{s}}
\newcommand{\be}{\begin{enumerate}}
\newcommand{\ee}{\end{enumerate}}
\newcommand{\beq}{\begin{equation}}
\newcommand{\eeq}{\end{equation}}

\newcommand{\npts}{\omega_{N}}
\newcommand{\nsd}{N^{1+s/d}}
\newcommand{\G}{{\mathcal G}_{s,d}}
\newcommand{\g}{g_{s,d}}
\newcommand{\underg}{\underline{g}_{s,d}}
\newcommand{\overg}{\overline{g}_{s,d}}
\newcommand{\undergd}{\underline{g}_{d,d}}
\newcommand{\overgd}{\overline{g}_{d,d}}
\newcommand{\gd}{g_{d,d}}
\newcommand{\Gd}{{\mathcal G}_{d,d}}
\newcommand{\End}{{\mathcal E}_{d}}
\newcommand{\Ed}{E_{d}}

\newcommand{\tsd}{\tau_{s,d}}
\newcommand{\tdd}{\tau_{d,d}}

\newtheorem{theorem}{Theorem}[section]
\newtheorem{lemma}[theorem]{Lemma}
\newtheorem{corollary}[theorem]{Corollary}
\newtheorem{proposition}[theorem]{Proposition}

\newtheorem{definition}{Definition}

\newcommand{\dist}{\text{\rm dist}}

\begin{document}

\begin{frontmatter}
    
    \title{Minimal Riesz energy point configurations for
    rectifiable $d$-dimensional manifolds}

\author{D.P. Hardin\corauthref{cor}} and 
\corauth[cor]{Corresponding author.}
\ead{doug.hardin@vanderbilt.edu}
\author{E.B. Saff\thanksref{ebsthanks}}
\thanks[ebsthanks]{The research of this author was supported, in part, 
by the U. S. National Science Foundation under grant DMS-0296026.}
\address{Department of Mathematics,
Vanderbilt University,
Nashville, TN 37240, USA}
\ead{esaff@math.vanderbilt.edu}


\begin{abstract}
    We investigate the energy of arrangements of $N$ points on a 
    rectifiable $d$-dimensional manifold $A\subset \R^{d'}$ that 
    interact  through the power law (Riesz) potential $V=1/r^{s}$, 
    where $s>0$ and $r$ is  Euclidean distance in $\R^{d'}$.  With 
    $\En(A,N)$ denoting the {\em minimal} energy  for such $N$-point 
    configurations, we determine the asymptotic behavior (as 
    $N\to\infty$) of  $\En(A,N)$ for each fixed $s\ge d$.  Moreover, 
    if $A$ has positive $d$-dimensional Hausdorff measure, we show 
    that $N$-point configurations  on $A$ that minimize the 
    $s$-energy are asymptotically uniformly distributed with respect 
    to $d$-dimensional Hausdorff measure on $A$ when $s\ge d$.  Even 
    for the unit sphere $S^{d}\subset \R^{d+1}$, these results are 
    new. 
    \end{abstract}

\end{frontmatter}

\section{Introduction}
Determining $N$ points on the unit sphere $S^{d}$ in $\R^{d+1}$ that 
are in some sense uniformly distributed over its surface is a 
classical problem that has applications to such diverse fields as 
crystallography, electrostatics, viral morphology, molecular 
modeling,  and global positioning.  Various criteria (appropriate to 
the application)  for the generation of such points  include 
best-packing, minimization of energy (e.g.,  Coulomb potentials), 
spherical $t$-designs (cubature), maximization of volume of convex 
polyhedra with $N$ vertices on $S^{d}$, etc.

A motivation for the present paper is the analysis of the asymptotic 
behavior (as $N\to\infty$) of optimal (and near optimal) $N$-point 
configurations that minimize the {\bf Riesz $s$-energy}
\beq \label{sEn}
\sum_{i\neq j}\frac{1}{|x_{i}-x_{j}|^{s}}
\eeq
over all $N$-point subsets $\{x_{1},\ldots,x_{N}\}$ of $S^{d}$, 
where $s >0 $ is a fixed parameter and $|\cdot|$ 
denotes the Euclidean norm in $\R^{d+1}$.  We remark that as $s\to 
\infty$, with $N$ fixed, the $s$-energy (\ref{sEn}) is increasingly 
dominated by the term(s) involving the smallest of pairwise distances 
and, in this sense, leads to the best-packing problem on $S^{d}$ (cf. 
\cite{CS}, \cite{Cox}).  We 
further note that for $s=1$ and $d=2$, the minimization of (\ref{sEn}) 
is the classic Thomson problem (see e.g. \cite{Bau}, \cite{Bow}, \cite{MKS}, \cite{Whyte}).

In this paper we investigate the case when $s$ is fixed, $s\ge d$, 
and $N\to \infty$.  Significantly our results apply not only to the 
sphere, but to a class of rectifiable $d$-dimensional manifolds 
embedded in $\R^{d'}$.  For such manifolds we determine, for $s\ge d$, 
the asymptotic behavior of the minimum Riesz $s$-energy as well as the 
asymptotic distribution of optimal and near optimal $N$-point 
configurations.  Indeed we shall prove that the latter is given by 
$d$-dimensional Hausdorff measure on the manifold and that the 
minimum $N$-point Riesz $s$-energy  over the manifold is 
asymptotically given by $C_{s}N^{1+s/d}$ when $s>d$ and by 
$C_{d}N^{2}\log N$ when $s=d$.  The essential feature of these 
results (see Theorems~\ref{main1}, \ref{main2}, and \ref{main3})  is not merely the order of growth 
of the minimum energy as $N\to \infty$, but rather the more delicate 
verification of the existence of the positive constants $C_{s}$ for 
$s\ge d$; a fact which is new even for the case of the sphere $S^{d}$ 
when $s>d$.  Somewhat surprising is the fact that we can give an 
explicit formula for $C_{d}$ (i.e., for the case $s=d$) in terms of 
its Hausdorff measure for any compact subset of a $d$-dimensional 
$C^1$-manifold in $\R^{d'}$ (see Theorem~\ref{main3} and 
equation~(\ref{Cdd})).

We remark that for $0 < s < d$, standard potential theoretic 
arguments can be used for the analysis of the minimum energy points 
(cf. \cite{Landkof}).  However, for $s\ge d$ such methods do not 
apply. Instead we exploit the scaling and translation properties of 
the energy function together with self-similarity and convexity 
arguments. 

For the remainder of this section we introduce some needed notation 
and, by way of further background, we mention known related results 
for the sphere $S^{d}$.  We devote the next section to the statement 
of our main results.  

Throughout this paper, $\npts=\{x_{1},\ldots,x_{N}\}$
denotes  a set of $N$ (possibly 0) distinct points in $\R^{d'}$.  For each real
$s>0$   the {\bf $s$-energy} of $\npts$ is 
given by 
\beq \label{Edef}
\E(\npts) := \sum_{x\neq y \in 
\npts}\frac{1}{|x-y|^{s}}=\sum_{y\in\npts}\sum_{\substack{x\in \npts 
\\ x\neq y}}\frac{1}{|x-y|^{s}}
\eeq
where, as above,  $|\cdot|$ 
denotes  Euclidean distance in $\R^{d'}$.  
For $A\subset \R^{d'}$  we define the {\bf $N$-point minimal 
$s$-energy over} $A$ by
\beq\label{EnDef}           
\En(A,N):=
\inf_{\npts \subset A} \E(\npts). 
\eeq
By convention, the  sum over an empty set of indices is taken to be zero and the infimum
over an empty set is $\infty$. Hence, 
$\En(A,N)=\infty$ if $N$ is greater than the cardinality of $A$ and 
$\E(\npts)=0$ if $N=0,1$.  It is clear that $\En(A,N)=\En({\bar 
A},N)$,  where ${\bar A}$ denotes the closure of $A$ and, furthermore, 
that $\En(A,N)=0$ if $A$ is unbounded.  Hence, without loss of 
generality, we may restrict ourselves to the case that $A$ is compact. 

For the unit sphere $S^{d}\subset \R^{d+1}$, the asymptotic behavior 
(as $N\to \infty$) of $\En(S^{d}, N)$ is quite different for the three 
cases ({\it i}) $0<s<d$; ({\it ii}) $s=d$; and ({\it iii}) $s>d$. The reason 
for this is that in case ({\it i}), the energy integral
\beq \label{pot}
I_{s}(\mu):= \iint_{S^{d}\times S^{d}}\frac{1}{|x-y|^{s}}\, d\mu(x)\, 
d\mu(y)
\eeq
taken over all probability measures $\mu$ supported on $S^{d}$ is 
{\em minimal} for normalized Lebesgue measure 
$\mc{H}_{d}(\cdot)|_{S^d}/\mc{H}_{d}(S^{d})$ on 
$S^{d}$.  However, for $s\ge d$, we have $I_{s}(\mu)=+\infty$ for all 
such measures $\mu$.  Roughly speaking, as the parameter $s$ 
increases, there is a transition from the domination of global effects 
to the domination of more local (near-neighbors) influences, and 
this transition occurs precisely when $s=d$.  

The following results are known for the above mentioned cases.  In 
case ({\it i}), classical potential theory yields (cf. \cite{Landkof}):
\begin{theorem} \label{known1} If $0<s<d$, 
    \beq
    \lim_{N\to \infty} 
    \frac{\En(S^{d},N)}{N^{2}}=
     I_{s}\left(\frac{\mc{H}_{d}(\cdot)|_{S^{d}}}{\mc{H}_{d}(S^{d})}\right),
    \eeq
    where $I_{s}$ is defined in (\ref{pot}).  Moreover,  any 
    sequence of optimal $s$-energy configurations 
    $(\npts^{*})_{2}^{\infty}\subset S^{d}$ is asymptotically 
    uniformly  distributed in the sense that for the weak-star 
    topology of measures, 
    \beq \label{UnifLim1}
    \frac{1}{N}\sum_{x\in\npts^{*}} \delta_{x} \longrightarrow 
    \frac{\mc{H}_{d}(\cdot)|_{S^{d}}}{\mc{H}_{d}(S^{d})} \qquad \text{ as 
    $N \to \infty$,}
    \eeq
    where $\delta_{x}$ denotes the unit point mass at $x$.
     \end{theorem}

    For case ({\it ii}), we have from the results of Kuijlaars and 
    Saff \cite{KS} and G\"otz and Saff \cite{GS} the following:
 \begin{theorem} \label{known2} Let $\mc{B}^{d}:={\bar B}(0,1)$ denote 
 closed unit ball in $\R^{d}$.   For $s=d$, 
     \beq \label{Sdlimit}
     \lim_{N\to\infty}\frac{\End(S^{d}, N)}{N^{2}\log  
     N}=\frac{\mc{H}_{d}(\mc{B}^{d})}{\mc{H}_{d}(S^{d})}=\frac{1}{d}\frac{\Gamma(\frac{d+1}{2})}{\sqrt{\pi}\Gamma(\frac{d}{2})},
     \eeq
     and  any sequence $(\npts^{*})\subset S^{d}$ of optimal $d$-energy 
      configurations satisfies (\ref{UnifLim1}).  
     \end{theorem}
        (The reader is cautioned that the definition of energy used here differs 
	   by a factor of 2 from 
	   that in \cite{KS}.)

     Until now, results for $s>d$ have been less complete, 
     describing only the order of growth of $\En(S^{d},N)$.  The 
     following is proved  in \cite{KS}.  
     
     \begin{theorem}\label{known3} For $s>d$, there exist positive 
     constants $c_{1}=c_{1}(s,d)$, $c_{2}=c_{2}(s,d)$ such that 
     $$
     c_{1}N^{1+s/d}\le \En(S^{d},N)\le c_{2}N^{1+s/d}, \qquad N\ge 2. 
     $$
     \end{theorem}
     
     Natural questions that therefore arise for the case $s>d$ are:
     \begin{enumerate}
	 \item[(a)] Does the limit 
	 $$
	 \lim_{N\to \infty}\frac{\En(S^{d},N)}{N^{1+s/d}}\quad  \text{ exist?}
	 $$
	 \item[(b)] If so, what is the limit?
	 \item[(c)] Are optimal $s$-energy configurations $\npts^{*}\subset 
	 S^{d}$ asymptotically uniformly distributed on $S^{d}$?  
	 \end{enumerate}
	 
	 In this paper we show as a corollary to our main results that 
	 questions (a) and (c) have affirmative answers.  Question (b) 
	 remains open for $d\ge 2$.  But more interesting is the fact that we 
	 can affirm (a) and (c) for a general class of $d$-dimensional 
	 rectifiable manifolds embedded in $\R^{d'}$ (cf. 
	 Theorem~\ref{main3}). 
	 And for such manifolds, in the case $s=d$, we give an explicit 
	 formula for $\lim_{N\to \infty}\End(A,N)/N^{2}\log N$ for every 
	 $d\in\N$.  For further background discussion, see \cite{HSnotices,RSZ1,RSZ2,SK}.
   
\section{Main Results}
In this section we state our main results.  Their proofs are given in 
the sections that follow. 
Let $\mc{H}_{d}$ denote $d$-dimensional Hausdorff measure in 
 $\R^{d'}$ normalized so that a $d$-sided cube with side length 1 has 
 $\mc{H}_{d}$-measure equal to 1.  In the case $d'=d$, then $\mc{H}_{d}$ reduces to Lebesgue 
 measure on $\R^{d}$.  
        \begin{theorem}\label{main1}
    Suppose $A\subset \R^{d}$ is compact.  Then 
    \beq \label{Cdd}
    \lim_{N\to 
    \infty}\frac{{\mc E}_{d}(A,N)}{N^{2}\log N}=
    \frac{\mc{H}_{d}(\mc{B}^{d})}{\mc{H}_{d}(A)},
   \eeq 
   where $\mc{B}^{d}$ is the closed unit ball in $\R^{d}$.  
   Furthermore,
    for $s>d$, the limit $\lim_{N\to \infty}\En(A,N)/N^{1+s/d}$ exists 
    and is given by
    \beq \label{Csd}
    \lim_{N\to 
    \infty}\frac{\En(A,N)}{N^{1+s/d}}=\frac{C_{s,d}}{\mc{H}_{d}(A)^{s/d}},
    \eeq
    where $C_{s,d}$ is a finite positive constant independent of $A$.  
    \end{theorem}

 \noindent  {\bf Remarks.} 
 \begin{enumerate}
  \item[(i)]  From (\ref{Csd}) it is clear that, for $s>d$, 
   \beq \label{CsdU}
    C_{s,d}= \lim_{N\to 
    \infty}\frac{\En(U^{d},N)}{N^{1+s/d}},
    \eeq
    where $U^{d}:=[0,1]^{d}$ is the unit cube in $\R^{d}$.  We further 
    remark that if $\mc{H}_{d}(A)=0$, then the limits in (\ref{Cdd}) and 
    (\ref{Csd}) equal $\infty$.  
 \item[(ii)]  Let $\bar B(a,\rho)$ denote the closed ball in $\R^{d}$ centered at $a$
 with radius $\rho$.  Then the  limit with $A=\bar B(a,\rho)$ in (\ref{Cdd}) 
 is simply $1/\rho^{d}$.   

   \end{enumerate}
    
        \begin{theorem} \label{main2}
   Let $A\subset \R^{d}$ be compact with $\mc{H}_{d}(A)>0$, and 
   $\npts=\{x_{k,N}\}_{k=1}^N$ be a sequence of asymptotically 
   optimal $N$-point  configurations in $A$ in the 
   sense that for some $s> d$ 
      \beq \label{Esd}
    \lim_{N\to 
    \infty}\frac{{E}_{s}(\npts)}{N^{1+s/d}}=
   \frac{C_{s,d}}{\mc{H}_{d}(A)^{s/d}},
   \eeq 
 or
    \beq \label{Edd}
    \lim_{N\to 
    \infty}\frac{{E}_{d}(\npts)}{N^{2}\log N}=
    \frac{\mc{H}_{d}(\mc{B}^{d})}{\mc{H}_{d}(A)}.
   \eeq 
   Let $\delta_{x}$ denote the unit point mass in the point $x$.  Then 
   in the weak-star topology of measures we have
   \beq \label{wconv}
   \frac{1}{N}\sum_{i=1}^{N}\delta_{x_{i,N}} \longrightarrow 
   \frac{\mc{H}_{d}(\cdot)|_{A}}{\mc{H}_{d}(A)} \quad \text{as $N\to \infty$.}
   \eeq
  \end{theorem} 
 
   \noindent  {\bf Remark.} 
 The convergence assertion (\ref{wconv}) is equivalent to  each of the 
 following assertions:
 \begin{enumerate}
     \item[(i)] For each $f$ continuous on $A$, 
    \beq 
    \lim_{N\to \infty}\frac{1}{N}\sum_{i=1}^{N}f(x_{i,N})=\frac{1}{\mc{H}_{d}(A)}\int_{A}f(x)\, d\mc{H}_{d}(x)
    \eeq
    \item[(ii)]  For every measurable set $B\subset A$ whose boundary 
    relative to $A$ has $\mc{H}_{d}$-measure zero,  the cardinality $|B\cap \npts|$ satisfies
   \beq \label{wconv2}
   \frac{|B\cap\npts|}{N}\to \frac{\mc{H}_{d}(B)}{\mc{H}_{d}(A)} \quad \text{ as 
   $N\to \infty$.}
    \eeq
 \end{enumerate}   
 
 \begin{theorem}\label{main4}
Let $A$ be a compact set in $\R^{d}$ 
such that $\mc{H}_{d}(A)>0$   and
$\lambda_{N}^{*}=\{x_{1,N}^{*},\ldots, x_{N,N}^{*}\}\subset A$ an 
optimal  $N$ point $s$-energy configuration for $A$.  If $s\ge d$, 
there exists a positive constant $C=C(A,s,d)$ such that for every $N\ge 
2$, 
\beq \label{sepcond}
\min_{i\neq j}|x_{i,N}^{*}-x_{j,N}^{*}|\ge \begin{cases} 
C/N^{1/d} & \text{for $s>d$,}\\
C/(N\log N)^{1/d} & \text{for $s=d$.}\end{cases}
\eeq
     \end{theorem}
  
     Recall that a mapping $\phi:T\to \R^{d'}$, $T\subset\R^{d}$,  is said to be a 
     {\bf Lipschitz mapping on $T$} if there is some constant $L$ such that 
     \beq \label{Lipdef}
   |\phi(x)-\phi(y)|\le L|x-y|\qquad \text{for $x,y\in T$}
    \eeq 
    and that $\phi$ is said to be a {\bf bi-Lipschitz mapping on $T$ 
    (with constant $L$)} if  
     \beq \label{biLipdef}
   (1/L)|x-y|\le|\phi(x)-\phi(y)|\le L|x-y|\qquad \text{for $x,y\in T$}.
    \eeq

 We say that $A\subset\R^{d'}$ is a {\bf $d$-rectifiable 
    manifold} if $A$ can be written as 
    \beq
    A=\bigcup_{k=1}^{n}\phi_{k}(K_{k})
    \eeq
    where, for each $k=1,\ldots, n$, $K_{k}\subset\R^{d}$ is compact and 
    $\phi_{k}$ is bi-Lipschitz 
    on an open set $G_k\supset K_{k}$.  
    Obviously any compact subset of a $d$-rectifiable 
    manifold is a $d$-rectifiable 
    manifold.    
    
% 
%     Note that a Lipschitz mapping $\phi$ on $T\subset\R^{d}$ may be extended 
%    to a Lipschitz mapping $\psi$ on $\R^{d}$ such that 
%    $\phi=\psi|_{T}$ and such that Lip($\phi$)=Lip($\psi$) (see 
%    \cite[2.10.43]{Fed}). 
%     
%     
%     

%      A set $A\subset \R^{d'}$ is said to be a {\bf $d$-dimensional rectifiable 
%      set} if $A$ is $\mc{H}_{d}$ measurable, $\mc{H}_{d}(A)<\infty$,
%      and $\mc{H}_{d}$-almost all of $A$ is contained in the countable 
%      union of Lipschitz images of bounded subsets of $\R^{d}$
%      (see \cite{Fed, Morgan}).
%      
     \begin{theorem}\label{main3}
	 Suppose $A\subset\R^{d'}$  is a $d$-rectifiable manifold and $s\ge d$.  If $s=d$, we further suppose that $A$ is a subset of a $d$-dimensional $C^1$-manifold.  
	 Then (\ref{Cdd}) 
	 and (\ref{Csd}) hold.  Furthermore, if $\mc{H}_{d}(A)>0$,  then 
	 (\ref{wconv}) holds for any asymptotically minimal sequence of $N$ 
	 point configurations $\npts$ for $A$ satisfying (\ref{Esd}) or  
	 (\ref{Edd}).  For the case when $A$ is  a bi-Lipschitz image of a single
	compact set in $\R^{d}$ and $\mc{H}_{d}(A)>0$, the separation estimates of (\ref{sepcond}) hold
	for any optimal $N$-point $s$-energy  configuration. 
	 \end{theorem}
	 
	    \noindent  {\bf Remark.} Note that $d'$ does not explicitly appear in (\ref{Cdd}) 
	 and (\ref{Csd}) but arises only in the norms for the computation of 
	 the energy.
	 
	   It is shown in \cite{KS} that, for the unit interval 
	   $U^{1}=[0,1]$, 
   \beq
    \lim_{N\to \infty}\frac{\En(U^{1},N)}{N^{1+s}}=2\zeta(s) \qquad 
    (s>1),
   \eeq
  where $\zeta(s)$ denotes the classical Riemann zeta function. 
  Hence, using (\ref{CsdU}), we get $C_{s,1}=2\zeta(s)$ for $s>1$.  
  Consequently, Theorem~\ref{main3} gives the following. 
    \begin{corollary} \label{cor1}
	Suppose $A$ is a compact subset of a $1$-rectifiable manifold in $\R^{d'}$ and $s>1$.  Then 
 \beq
    \lim_{N\to 
    \infty}\frac{\En(A,N)}{N^{1+s}}=\frac{2\zeta(s)}{\mc{H}_{1}(A)^{s}}.
   \eeq	
	\end{corollary}
    That Corollary~\ref{cor1} holds 
     when $A$ is a finite union of rectifiable 
     Jordan arcs was shown in \cite{MMRS}.  Since a Lipschitz mapping on 
     an interval is absolutely continuous, but the converse is not 
     necessarily true, the results in \cite{MMRS} hold  in cases not 
     covered by
     Corollary~\ref{cor1}.  On the other hand,
     Corollary~\ref{cor1} 
     applies to 1-rectifiable manifolds 
     that are not covered by the results in \cite{MMRS} such as, for example, 
     when $A$ is the bi-Lipschitz  image of a Cantor subset of 
     [0,1] having positive measure.  
     
   For the $2$-sphere it is shown in \cite{KS} that for $s>2$,
     \beq \label{Cs2}
     \limsup_{N\to  
     \infty}\frac{\En(S^{2},N)}{N^{1+s/2}}\le\left(\frac{\sqrt{3}}{8\pi}\right)^{s/2}
     \zeta_{L}(s),
    \eeq
     where $\zeta_{L}(s)$ is the zeta function for the hexagonal 
     lattice $L$ consisting of points of the form 
     $m(1,0)+n (1/2,\sqrt{3}/2)$ for  $m,n\in\Z$.  
     Consequently (cf. (\ref{Csd})), 
     \beq
     \label{Cs22}
     C_{s,2}\le\left(\frac{\sqrt{3}}{2}\right)^{s/2}\zeta_{L}(s).\eeq
     It is conjectured in \cite{KS} that equality holds in  (\ref{Cs2}) 
     which, if true, would imply that equality holds in (\ref{Cs22}).

     An outline of the remainder of the paper is as follows. In 
     Section 3 we establish some basic lemmas on the minimal 
     $s$-energy of the union of two subsets of $\R^{d}$.  Section 4 
     gives the proof of Theorem~\ref{main1} for the special case when 
     $A$ is the unit cube in $\R^{d}$.  In Section 5, we verify 
     Theorems~\ref{main1} and \ref{main2} for almost clopen sets in 
     $\R^{d}$.  Results on the separation of points in optimal energy 
     configurations are established in Section 6. The proofs of 
     Theorems~\ref{main1} and \ref{main2} for general compact sets 
     in $\R^{d}$ is presented in Section 7 and the proof of 
     Theorem~\ref{main3} appears in Section 8.

\section{Basic Lemmas}
In this section we establish several lemmas that are required for the 
proofs of our main results. 
First we establish that if  $A\subset \R^{d}$ is bounded with 
nonempty interior, then $\En(A,N)$ grows as $N\to \infty$ with order $N^{1+s/d}$ for $s>d$ and 
 $N^{2}\log N$ for $s=d$.

% \begin{lemma}\label{cs}
%     Suppose $a_{i}>0$ for $i=1,\ldots,N$. Then
%     $$
%     \sum_{i=1}^{N}a_{i}\sum_{i=1}^{N}\frac{1}{a_{i}}\ge N^{2}
%     $$
%     \end{lemma}
%  	
% \begin{lemma}[Jensen or H\"{o}lder Inequality]\label{JenIn}
%     Suppose $a_{i}\ge 0$ for $i=1,\ldots,M$ and $p> 1$. Then
%  $$\frac{1}{M}\sum_{i=1}^{M} a_{i}^{p}\ge 
%  \left(\frac{1}{M}\sum_{i=1}^{M} a_{i}\right)^{p} $$
%  with equality if and only if $a_{i}=a_{j}$ for all $i,j=1,\ldots, M$.
%  If  $\sum a_{i}=1$ and $\delta>0$, then there is some 
%  $\epsilon=\epsilon(\delta,M,p)>0$ such that $\sum_{i=1}^{M}a_{i}^{p}>M^{1-p}+\epsilon$ whenever
%  $\ds \max_{i}|a_{i}-1/M|>\delta$. 
% \end{lemma}
	
\begin{lemma}\label{C0C1}
    Suppose $A\subset \R^{d}$ is a bounded set with nonempty interior.  
    There exist positive constants 
    $C_{0},\,C_{1}$  (depending on $A$, $s$, and $d$, but not on $N$)
    such that, if $s>d$,
   \beq
       \label{bnd1}
       C_{0}\nsd \le \En(A,N) \le C_{1}\nsd \qquad (N\ge 2) 
       \eeq
    and, if $s=d$, 
     \beq
       \label{bnd2}
       C_{0}N^{2}\log N \le \mc{E}_{d}(A,N) \le C_{1}N^{2}\log N \qquad (N\ge 2).
        \eeq
   \end{lemma}
   
   \beginproof
              We first consider $U^{d}=[0,1]^{d}$. Let $B(x,r)$ 
              denote the open ball in $\R^{d}$ with center $x$ and 
              radius $r$.  Then with $C=(1/2)^{d}\mc{H}_{d}(B(0,1))$ we have
       \beq \label{mULB}
	   \mc{H}_{d}(B(x,r)\cap U^{d})\ge Cr^{d}
	   \eeq
	   for any $x\in U^{d}$ and  $r<1$. 
        
       For $N>1$, let $\npts=\{x_{1},\ldots,x_{N}\}$ be a collection of 
       $N$ distinct points in $U^{d}$ and let $$r_{i}:=\min_{j\neq i} |x_{i}-x_{j}|.$$ Since 
       $B(x_{i},r_{i}/2)\cap B(x_{j},r_{j}/2)=\emptyset$ for $1\le 
       i\neq j \le N$, we have
       \beq\label{mALB}
	  1= \mc{H}_{d}(U^{d})\ge \sum_{i=1}^{N}\mc{H}_{d}(B(x_{i},r_{i}/2)\cap U^{d})\ge 
       \frac{C}{2^{d}}\sum_{i=1}^{N}r_{i}^{d}.
       \eeq
      By the Cauchy-Schwarz inequality we have 
      \beq
      \label{csin}
      N^{2}=\left(\sum_{i=1}^{N}r_{i}^{d/2}r_{i}^{-d/2}\right)^{2}\le\sum_{i=1}^{N}r_{i}^{d}
      \sum_{i=1}^{N}r_{i}^{-d},
      \eeq
      which is known as the harmonic-arithmetic mean inequality.  
      Thus
      \begin{align} \label{nptsLB}
	   \E(\npts) &\ge \sum_{i=1}^{N}\frac{1}{r_{i}^{s}} 
	   =N\sum_{i=1}^{N}\frac{1}{N}\left(\frac{1}{r_{i}^{d}}\right)^{s/d} \\  
	   &\ge N\left(\sum_{i=1}^{N}\frac{1}{N}\frac{1}{r_{i}^{d}}\right)^{s/d} 
	\ge N\left(\frac{N}{\sum_{i=1}^{N}r_{i}^{d}}\right)^{s/d}, \nonumber
       \end{align}
       where the next to the last inequality follows from Jensen's 
       inequality (or H\"{o}lder's inequality) and the last 
       inequality follows from (\ref{csin}). Since (\ref{nptsLB}) 
       holds for any collection $\npts$ of $N$ distinct points  in 
       $U^{d}$, then using (\ref{mALB}) we have
       $$
       \En(U^{d},N)\ge \nsd C^{s/d}2^{-s} \qquad (N>2)
       $$
       showing that the lower estimate in (\ref{bnd1}) holds for $A=U^{d}$ and with $C_{0}=C^{s/d}2^{-s}$.  
       
       For $s=d$, the lower estimate in (\ref{bnd2}) is not so 
       straightforward. For this case we shall apply the known 
       result (\ref{Sdlimit}).
	 The unit cube $U^{d}$ in $\R^{d}$ can be 
	   projected onto a subset of $S^{d}$ via the stereographic 
	   projection $\sP:\R^{d}\to S^{d}$ defined by 
	   \beq \label{sPdef}
	   \sP(x)=(tx,1-t)\in \R^{d+1}, \qquad t=\frac{2}{|x|^{2}+1}.
	   \eeq
	   It is easily verified (and well-known in the case $d=2$) that for
	   $x,y\in\R^{d}$, we have 
	   \beq \label{sPdiff}
	   |\sP(x)-\sP(y)|=\frac{2|x-y|}{\sqrt{1+|x|^{2}}\sqrt{1+|y|^{2}}}.
	  \eeq
	   Consequently, for some positive constant $C$, 
	   $$\mc{E}_{d}(U^{d},N)\ge C\mc{E}_{d}(\sP(U^{d}),N)\ge 
	   C\mc{E}_{d}(S^{d},N),
	   $$
	   and so the desired lower estimate follows from (\ref{Sdlimit}). 
	   (Later we shall show how (\ref{Sdlimit}) can be utilized to 
	   determine the precise asymptotic behavior of 
	   $\mc{E}_{d}(A,N)$ for $d$-rectifiable manifolds.)
       
       For $N>1$, let $m$ be the positive integer such that 
       $m^{d}\le N < (m+1)^{d}$.  
       Let $\npts$ consist of $N$ points selected from $U^{d}\cap 
       \left(\Z^{d}/m\right)$. 
       Then $x-y\in\Z^{d}/m$ for $x,y\in\npts$.  For 
       ${\bf k}=(k_{1},\ldots,k_{d})\in\Z^{d}$, set 
       $\|{\bf k}\|_{\infty}=\max\{|k_{i}|,i=1,\ldots,d\}$ and let 
       $|{\bf k}|$ 
       denote its Euclidean norm.  Then
  \begin{align*}
      \E(\npts)&=\sum_{x\neq y\in\npts}\frac{1}{|x-y|^{s}} \le 
      N m^{s}\sum_{1\le \|{\bf k}\|_{\infty}\le m}\frac{1}{|{\bf k}|^{s}}\\ % corr 1
    &\le \nsd  \sum_{j=1}^{m}\sum_{\|{\bf k}\|_{\infty}=j}\frac{1}{j^{s}}
      =\nsd\sum_{j=1}^{m}\frac{(2j+1)^{d}-(2j-1)^{d}}{j^{s}}\\
     &\le C \nsd \sum_{j=1}^{[N^{1/d}]+1}\frac{1}{j^{1+s-d}}.
      \end{align*}
     For $s>d$, the sum in the last inequality is bounded from above 
     independently of $N$, while for $s=d$, it is bounded by a 
     constant times $\log N$. Thus the estimates (\ref{bnd1}) and 
     (\ref{bnd2}) hold for $A=U^{d}$.  % restrict sum over finite set.
      
    More generally, if $A$ is a bounded set with nonempty interior, then there 
      exist $r,R>0$ and $x_{0},x_{1}\in\R^{d}$ such that 
      $r U^{d}+x_{0}\subset A\subset R U^{d}+x_{1}$. Since
      $\En(\rho U^{d}+x, N)=\rho^{-s}\En(U^{d}, N)$ for any $\rho>0$ and $x\in 
      \R^{d}$, the estimates (\ref{bnd1}) and (\ref{bnd2}) follow for $A$. 
   (For an alternative proof of the upper bound, see the proof of  Theorem~\ref{main4} in Section 6.)
       \eproof

\begin{definition} Let $\tsd:\N\to \R$  be defined by
\beq \tsd(N)=  
 \begin{cases}
 \nsd & \text{ if $s>d$}\\
 N^{2}\log N & \text{ if $s=d$}.
      \end{cases} 
      \eeq
      For  $A\subset \R^{d'}$ and a positive integer $N$, 
     define 
  \beq
	 \G(A,N):=\En(A,N)/\tsd(N) 
	 \eeq
and let
     $$ \underg(A):=\liminf_{N\to \infty}\G(A,N), \quad
     \overg(A):=\limsup_{N\to \infty}\G(A,N).$$
     We set $$\g(A):=\lim_{N\to \infty}\G(A,N)$$ when the limit (as an 
     extended real number)
     exists. 
  \end{definition}   
     
     If $A\subset\R^{d}$ is bounded and  has nonempty interior, then by Lemma~\ref{C0C1} there 
     exist positive constants $C_{0},C_{1}$ such that 
     $$C_{0}\le \G(A,N)\le C_{1} $$ 
     for $N\ge 2$. Hence,
     $\underg(A)$ and $\overg(A)$ are both positive and 
     finite in this case.

     \begin{lemma}\label{unionlemmalower}
	 Suppose $s\ge d$ and that $A$ and $B$ are bounded sets in $\R^{d'}$.  
Then \beq\label{glower} \underg(A\cup B)\ge 
	    \left(\underg(A)^{-d/s}+\underg(B)^{-d/s}\right)^{-s/d}.
	    \eeq
	    Furthermore, if $\underg(A)<\infty$ or if $\underg(A)=\infty$ 
	    and $\underg(B)<\infty$, and $\mc{N}$ is an infinite subset of $\N$ and 
 $(\npts)_{N\in\mc{N}}$ is a sequence of sets   
 $\npts\subset A\cup B$, $N\in\mc{N}$, such that  
\beq \label{glowerlimitcond}
    \lim_{\substack{N\to\infty\\ 
    N\in\mc{N}}}\frac{\E(\npts)}{\tsd(N)}= 
    \left(\underg(A)^{-d/s}+\underg(B)^{-d/s}\right)^{-s/d},
\eeq
then  
\beq \label{glowerlimit}
    \lim_{\substack{N\to\infty \\ N\in\mc{N}}}\frac{|\npts\cap A|}{N} =
\frac{\underg(B)^{d/s}}{\underg(A)^{d/s}+\underg(B)^{d/s}}.
\eeq
\end{lemma}
     
     {\noindent \bf Remark.} If  both
     $\underg(A)<\infty$ and $\underg(B)=\infty$, then the right-hand sides of (\ref{glower})
     and (\ref{glowerlimit})
     are understood to be $\underg(A)$ and 1, respectively; while if 
     $\underg(A)=\underg(B)=\infty$,
     then the right-hand side of (\ref{glower}) is understood 
     to be $\infty$. 
     
\beginproof  Both $\underg(A)$ and $\underg(B)$ are positive since $A$ and $B$ are bounded.
    First we assume that $\underg(A)$ and $\underg(B)$ are finite. 
Suppose, for $N\in\N$, that $\npts$ is a set of $N$ distinct points in 
$A\cup B$.  Let $\npts^{A} =\npts\cap A$ and 
$\npts^{B}=\npts\setminus\npts^{A}$.  Then  
\beq\label{glowereq0}
    \E(\npts)=\E(\npts^{A})+\E(\npts^{B})+
  2  \sum_{a\in\npts^{A},b\in\npts^{B}}\frac{1}{|a-b|^{s}}\ge \E(\npts^{A})+\E(\npts^{B})
\eeq
and  hence  
\beq \label{glowereq5}
\En(A\cup B,N)\ge 
\min_{N_{A}+N_{B}=N}\left(\En(A,N_{A})+\En(B,N_{B})\right),
\eeq
where $N_{A}$ and $N_{B}$ are nonnegative integers.  
First suppose $s>d$. Then we have
\begin{align}
\underg&(A\cup B) \nonumber \\ &\ge 
\liminf_{N\to \infty} 
\min_{N_{A}+N_{B}=N}\left[\frac{\En(A,N_{A})}{N_{A}^{1+s/d}}\left(\frac{N_{A}}{N}\right)^{1+s/d}+
\frac{\En(B,N_{B})}{N_{B}^{1+s/d}}\left(\frac{N_{B}}{N}\right)^{1+s/d} 
\right]\nonumber
\\ &\ge \label{glowereq1}
\liminf_{N\to \infty} \min_{N_{A}+N_{B}=N}\left[\underg(A)\left(\frac{N_{A}}{N}\right)^{1+s/d}+
\underg(B)\left(\frac{N_{B}}{N}\right)^{1+s/d} \right]
\\&\ge\min_{0\le \alpha\le 1}\left[\underg(A)\alpha^{1+s/d}+
\underg(B)\left(1-\alpha\right)^{1+s/d}\right].\nonumber
\end{align}
(Note: In the case $N_{A}=0$ we set $\frac{\En(A,N_{A})}{N_{A}^{1+s/d}}\left(\frac{N_{A}}{N}\right)^{1+s/d}
=\frac{\En(A,N_{A})}{\nsd}=0$. The case
$N_{B}=0$ is handled similarly.) 
Let 
\beq\label{glowereq2}
F(\alpha):=\underg(A)\alpha^{1+s/d}+
\underg(B)\left(1-\alpha\right)^{1+s/d} \qquad (0\le \alpha\le 1) .
\eeq
The reader may verify using elementary calculus that $F$ has a unique 
minimum value $F(\alpha^{*})=\left( 
\underg(A)^{-d/s}+\underg(B)^{-d/s}\right)^{-s/d}$,
where 
$$\alpha^{*}={\underg(B)^{d/s}}/{\left(\underg(A)^{d/s}+\underg(B)^{d/s}\right)}.$$
This proves (\ref{glower}) when $s>d$.  

Now suppose $(\npts)_{N\in\mc{N}}$ is a sequence of sets   
 $\npts\subset A\cup B$, $N\in\mc{N}$, such that  
(\ref{glowerlimitcond}) holds.  We may rewrite (\ref{glowereq0}) in the form
\beq \label{glowereq3}
\frac{\E(\npts)}{\nsd}\ge
\frac{\E(\npts^{A})}{N_{A}^{1+s/d}}\left(\frac{N_{A}}{N}\right)^{1+s/d}+
\frac{\E(\npts^{B})}{N_{B}^{1+s/d}}\left(\frac{N_{B}}{N}\right)^{1+s/d} 
\qquad (N\in \mc{N}),
\eeq
and hence, if $\beta$ is any limit point of the sequence
$N_{A}/N$, $N\in\mc{N}$, we get from (\ref{glowerlimitcond}) that 
$F(\alpha^{*})\ge F(\beta)$. Consequently, $\beta=\alpha^{*}$, which 
is equivalent to (\ref{glowerlimit}) in the case $s>d$.

We leave to the reader the remaining cases where at least one of
$\underg(A)$ and $\underg(B)$ is infinite.  It is helpful to regard 
separately the cases when $N_{B}$ remains bounded or when $N_{B}\to 
 \infty$ as $N\to \infty$.  
 
% Now suppose $\underg(B)=\infty$ and $\underg(A)<\infty$.  Let 
% $N_{A}^{*}$ and $N_{B}^{*}$ be values of $N_{A}$ and $N_{B}$, 
% respectively for which (\ref{glowereq5}) is minimized.  If for some 
% subsequence of $N$, $N_{B}^{*}$ remains bounded then (\ref{glowereq1})
% shows that $\underg(A\cup B) \ge \underg(A)$.  On the other hand, if  
% a subsequence of $N_{B}^{*}$ goes to $\infty$,  then 
% one may replace $\underg(B)$ in (\ref{glowereq1}) and (\ref{glowereq2})
% as well as
% ${\E(\npts^{B})}/{N_{B}^{1+s/d}}$ in (\ref{glowereq3})
% by an arbitrary positive 
% number $\gamma$. It then follows that $\underg(A\cup B)\ge 
% (\underg(A)^{-d/s}+\gamma^{-d/s})^{-s/d}$ and that any limit 
% point of the sequence $N_{A}/N$  is bounded below 
% by $\gamma^{d/s}/(\underg(A)^{d/s}+\gamma^{d/s})$.  Taking $\gamma\to 
% \infty$ shows  (\ref{glower}) and (\ref{glowerlimit}) hold in this 
% case.  Similarly,  if $\underg(A)=\underg(B)=\infty$, then one may 
% also replace $\underg(A)$ in (\ref{glowereq1}) and (\ref{glowereq2})
% (\ref{glowereq1}) and (\ref{glowereq2}) with an arbitrary positive constant 
% and which shows $\underg(A\cup B)=\infty$.
% 

 The 
 $s=d$ case of both (\ref{glower})  and (\ref{glowerlimit}) follows in a  similar 
 manner  and is left as well for the 
 reader.  
 \eproof

For $A,B\subset\R^{d'}$, let $\dist(A,B):=\inf_{a\in A,\, b\in 
B}|a-b|$.\\

\begin{lemma} \label{unionlemmaupper}
   Suppose $A$ and $B$ are bounded sets in $\R^{d'}$ such that
   $\dist(A,B)>0$. Then 
	 \beq\label{gupper} 
	 \overg(A\cup B)\le \left(\overg(A)^{-d/s}+\overg(B)^{-d/s}\right)^{-s/d}.
	\eeq
\end{lemma}

\beginproof
    If $\overg(A)$ or $\overg(B)$ equal zero then $\overg(A\cup B)=0$ 
    and so (\ref{gupper}) holds
    (note that the right hand side of (\ref{gupper}) is understood to 
    be zero in this case).  
    
   Now suppose  $\overg(A)$ and $\overg(B)$ are both positive. 
Let $\delta=\dist(A,B)$ and suppose $N\in \N$.  Let 
$N_{A}=[\alpha^{*}N]$ where 
$\alpha^{*}={\overg(B)^{d/s}}/\left({\overg(A)^{d/s}+\overg(B)^{d/s}}\right)$ and let 
$N_{B}=N-N_{A}$.  Then
 \begin{align*}
     \En(A\cup B,N) &\le 
     \En(A,N_{A})+\En(B,N_{B})+2\delta^{-s}N_{A}N_{B}\\
     &\le \En(A,N_{A})+\En(B,N_{B})+2\delta^{-s}N^{2},
 \end{align*}
and hence 
%\beq
\begin{align*}
     \G(A\cup B,N) 
     &\le \frac{\En(A,N_{A})}{\tsd(N_{A})}\frac{\tsd(N_{A})}{\tsd(N)}+
     \frac{\En(B,N_{B})}{\tsd(N_{B})}\frac{\tsd(N_{B})}{\tsd(N)}+2\delta^{-s}\frac{N^{2}}{\tsd(N)}.
 \end{align*}
 %\eeq

Observe that
$\lim_{N\to\infty}{\tsd(N_{A})}/{\tsd(N)}=(\alpha^{*})^{1+s/d}$, 
$\lim_{N\to\infty}{\tsd(N_{B})}/{\tsd(N)}=(1-\alpha^{*})^{1+s/d}$, 
and $\lim_{N\to\infty}N^{2}/\tsd(N)=0$.  Thus we have
\begin{align*}
\overg(A\cup B)&\le \overg(A)(\alpha^{*})^{1+s/d}+\overg(B)(1-\alpha^{*})^{1+s/d}\\
&=\left(\overg(A)^{-d/s}+\overg(B)^{-d/s}\right)^{-s/d}.
\end{align*}
\eproof

We say that a set $A\subset\R^{d}$ is {\bf scalable} if $A$ is closed 
and if for each 
$\epsilon>0$ there is some bi-Lipschitz mapping $h:A\to A^{\circ}$ with constant 
$(1+\epsilon)$ where $A^{\circ}$ denotes the interior of $A$. 
For example, a compact, convex set with nonempty 
interior is scalable since, for $\epsilon>0$,  
one may choose $h(x)=(1+\epsilon)^{-1}(x+\epsilon u)$ for any fixed $u$ in 
the interior of $A$.  Similarly, a star-like set is scalable.  \\

\begin{corollary} \label{AcupBCorollary}
    Suppose $s\ge d$ and $A$ and $B$ are  compact subsets of $\R^{d}$ 
    with disjoint interiors such %corr 2
    that $\g(A)$ and $\g(B)$ both exist and  $A$ is scalable. Then $\g(A\cup B)$ exists and 
    \beq\label{AcupB}
	\g(A\cup B)=
	    \left(\g(A)^{-d/s}+\g(B)^{-d/s}\right)^{-s/d}.
    \eeq
\end{corollary}
 
\beginproof
	Let  $0<\epsilon<1$. Since $A$ is 
	scalable, there is some bi-Lipschitz mapping $h$ with 
	constant $(1+\epsilon)$ such that
	$h(A)\subset A^{\circ}$ and, hence, $\dist(h(A),B)\ge 
	\dist(h(A),A^{c})>0$.  Then 
	Lemmas~\ref{unionlemmalower} and \ref{unionlemmaupper} imply
\begin{align*}
    \left(\g(A)^{-d/s}+\g(B)^{-d/s}\right)^{-s/d} &\le   \underg(A\cup B)
\\ &\le \overg(A\cup B) 
\le \overg(h(A)\cup B)
\\
&\le \left(\overg(h(A))^{-d/s}+\g(B)^{-d/s }\right)^{-s/d}.
\end{align*}
Since $\overg(h(A))\le (1+\epsilon)^{s}\g(A)$, on letting 
$\epsilon\to 0$, we get (\ref{AcupB}).
\eproof

     \section{The unit cube $U^{d}:=[0,1]^{d}$.}
     In this section we prove that $\g(U^{d})$ exists when $s\ge d$. We 
     first prove the result when $s>d$ by using the self-similarity 
     of $U^{d}$ to obtain estimates relating $\G(U^{d},N)$ at different values 
     of $N$. For $s=d$ the method is not immediately applicable.  Instead 
     we use results 
     and techniques developed in \cite{KS} and \cite{GS} for the sphere $S^{d}$ 
     which actually yield  $\gd(U^{d})$ explicitly.   The proof of the 
     next theorem in the case $s=d$ is given separately in Section~\ref{sedProof}.
     
 \begin{theorem} \label{gU} For $s\ge d$,
the limit  $\g(U^{d}):=\lim_{N\to \infty}\G(U^{d},N)$ 
	exists and is finite and positive. Moreover, in the case $s=d$, 
	\begin{equation}
	    \gd(U^{d})=\mc{H}_{d}(\mc{B}^{d})=\frac{2 
	    \pi^{d/2}}{d\,\Gamma(\frac{d}{2})}.
	    \end{equation}
\end{theorem}

  \beginproof
      {\sc Case: $s>d$.}\\   
      We first establish a lemma relating $\G(U^{d},N)$ at different 
      values of $N$. 
\begin{lemma}\label{UupBnd} Suppose $s> d$, $\gamma\in (0,1)$ 
    and $m$ is a positive integer. Then 
there is some constant $C>0$ 
(independent of $m$, $N$ or $\gamma$) such 
that
   \beq 
       \label{UupBndeq}
       \G(U^{d},m^{d}N)\le     \gamma^{-s}\G(U^{d},N)+C(1-\gamma)^{-s}N^{1-s/d}.
   \eeq
\end{lemma}

\beginproof  
    If $m$ is a positive integer,  let 
  $I_{m}=\{0,\ldots,m-1\}^{d}\subset \Z^{d}$  and, for $\bi\in I_{m}$, set
  $U_{m,\bi}:=(U^{d}+\bi)/m.$
   Note that $\G(U_{m,\bi},N)=m^{s}\G(U^{d},N)$.  

    For $\bi\in I_{m}$, let $\npts^{\bi}$ be a 
    set of $N$ points in $(\gamma U^{d}+\bi)/m$ with minimum energy.
   Let
    $\omega_{m^{d}N}=\bigcup_{\bi\in I_{m}}\npts^{\bi}$. 
    If $x\in\npts^{\bi}$ and $y\in\npts^{\bj}$, then 
    $|x-y|\ge \delta:=(1-\gamma)/m$ if $\|\bi-\bj\|_\infty=1$ and $|x-y|\ge 
   \|\bi-\bj\|_\infty/(2m)$ for $\|\bi-\bj\|_\infty>1$.   Then
\begin{align*}\label{EnEst1}
    \En(U^{d},&m^{d}N)\le \E(\omega_{m^{d}N})   
    \le \sum_{\bi\in I_{m}}\E(\npts^{\bi})+\sum_{\substack{\bi\neq \bj\in I_{m}\\ 
    x\in\npts^{\bi},\, y\in\npts^{\bj}}} \frac{1}{|x-y|^{s}} \\
    &\le \sum_{\bi\in I_{m}}\left(\En(\frac{\gamma}{m}U^{d},N) 
    +\delta^{-s}3^d N^{2}+2^{s}\sum_{\substack{\bj\in I_{m}\\ \|\bi-\bj\|_\infty>1}}m^{s} 
    \|\bi-\bj\|_\infty^{-s}N^{2}\right)\\
     &= \sum_{\bi\in I_{m}}\left(m^{s}\gamma^{-s}\En(U^{d},N) 
    +m^{s}\frac{3^d N^{2}}{(1-\gamma)^{s}}+2^{s}\sum_{\substack{\bj\in I_{m}\\ \|\bi-\bj\|_\infty>1}}m^{s} 
    \|\bi-\bj\|_\infty^{-s}N^{2}\right)\\
    &\le    
    m^{d+s}\left(\gamma^{-s}\En(U^{d},N)+\frac{3^d N^{2}}{(1-\gamma)^{s}}+2^{s}KN^{2}\right)
\end{align*}
where
$K:= \sum_{\mathbf{k}\in \Z^{d}\setminus \{\mathbf{0}\}} \|\mathbf{k}\|_{\infty}^{-s}$ is finite, and so
\beq\label{ineq0}
  \G(U^{d},m^{d} N)\le 
  \gamma^{-s}\G(U^{d},N)+(1-\gamma)^{-s}(3^{d}+2^{s}K)\frac{m^{s+d}N^{2}}{\tsd(m^{d}N)}.
\eeq
Since 
$m^{s+d}N^{2}/\tsd(m^{d}N)= 
N^{1-s/d}$,
 the inequality  (\ref{UupBndeq}) follows from (\ref{ineq0})  with 
    $C=3^{d}+2^{s}K$, which completes the proof of Lemma~\ref{UupBnd}. 
\eproof

Now suppose $\epsilon>0$ and $0<\gamma<1$. Let $C$ be the constant in 
	 (\ref{UupBndeq}) and  let $N^{*}$ be such that 
	 $\G(U^{d},N^{*})<\underg(U^{d})+\gamma^{s}\epsilon/2$ and 
	 $C(N^{*})^{1-s/d}<(1-\gamma)^{s}\epsilon/2$.
	  By Lemma~\ref{UupBnd} 
	 we then have 
	 $$
	 g_{m}:=\G(U^{d},m^{d}N^{*})<\gamma^{-s}(\underg(U^{d})+\gamma^{s}\epsilon/2)+C(1-\gamma)^{-s}(N^{*})^{1-s/d}
	 $$
	 for any $m\in\N$, and  hence
	 $$
	 \limsup_{m\to \infty} g_{m} \le \gamma^{-s} 
	 \underg(U^{d}) + \epsilon.
	 $$
	 
  For $N>N^{*}$, let $m_{N}$ be the greatest integer such that $m_{N}^{d} 
  N^{*} < N$.  Then  
	 $$\G(U^{d},N)=\frac{\En(U^{d},N)}{N^{1+s/d}}\le 
	 \frac{\En(U^{d},(m_{N}+1)^{d}N^{*})}{(m_{N}^{d}N^{*})^{1+s/d}}=(1+1/m_{N})^{s+d}g_{m_{N}+1} 
	 $$
	 holds for all $N>N^{*}$,
	 and thus $$\limsup_{N\to\infty}\G(U^{d},N)\le  
	 \limsup_{N\to\infty}(1+1/m_{N})^{s+d}g_{m_{N}+1}\le  \gamma^{-s} 
	 \underg(U^{d}) +\epsilon.$$
	 Since this holds for all $0<\gamma<1$ and $\epsilon>0$ we have that 
	 $\overg(U^{d})\le \underg(U^{d})$ and, hence, $\g(U^{d})$ exists.  By 
	 Lemma~\ref{C0C1}, $\g(U^{d})$ is finite and positive in the case $s>d$, 
	 which completes the proof in this case.  
	 \eproof

	\subsection{Proof of Theorem~\ref{gU} in the case $s=d$.}  \label{sedProof}
	 
	  By Theorem~\ref{known2} we know that  
	 $\gd(S^{d})$ exists and is given as in (\ref{Sdlimit}).
  For $N\in\N$, 
	 let $\npts^{*}$ denote a 
	 set of $N$  points in $S^{d}$ minimizing the $d$-energy.  Also from 
	 this theorem we have 
	\beq \label{Sduniformdist}
	 \lim_{N\to \infty}\frac{\vert \npts^{*}\cap A\vert}{N} = 
	 \frac{\mc{H}_{d}(A)}{\mc{H}_{d}(S^{d})}
	\eeq
	 whenever $A\subset S^{d}$ is such that the boundary $\partial A$
	 (relative to the sphere) has $\mc{H}_{d}(\partial A)=0$.  Such a set
	$A$ is called an {\bf almost clopen} subset of $S^{d}$.

	 \begin{lemma} \label{uniformenergylemma} For $N\in\N$, 
	 let $\npts^{*}$ denote a 
	 set of $N$  points in $S^{d}$ minimizing the $d$-energy. If $A$ 
	 is an almost clopen subset of $S^{d}$,  then
	\begin{equation} \label{uniformenergy}
	     \lim_{N\to \infty}\frac{E_{d}(\npts^{*}\cap 
	     A)}{\tdd(N)}=\gd(S^{d})\frac{\mc{H}_{d}(A)}{\mc{H}_{d}(S^{d})}.
	\end{equation}

	     \end{lemma}

	  \beginproof
	      We first show that for any almost clopen subset $K$ of $S^{d}$ 
	      we have 
	      \beq \label{UEeq1}
	      \limsup_{N\to \infty} \frac{E_{d}(\npts^{*}\cap K)}{\tdd(N)}
\le \gd(S^{d})\frac{\mc{H}_{d}(K)}{\mc{H}_{d}(S^{d})}.
\eeq
For this purpose we follow the argument given in \cite{KS}. Let 
$\{x_{i,N}^{*}\}_{i=1}^{N}$ denote the points of $\npts^{*}$ and for 
each $i$, set 
$$
U_{i,N}(x):=\sum_{j\neq i}\vert x-x_{j,N}^{*}\vert ^{-d} ,\qquad x\in S^{d}.
$$
It is shown in inequality (6.6) of \cite{KS} that for every $r>0$ 
sufficiently small we have 
\beq
\label{UEeq2} U_{i,N}(x^{*}_{i,N})\le \frac{\gd(S^{d})N\log 
N}{\left(1-r^{d}\gd(S^{d})\right)}  + \mc{O}_r (N)\qquad (N\to \infty). 
\eeq
Let $\Lambda(K,N):=\{i\vert x_{i,N}^{*}\in K\}$ and 
$N^{K}:=|\Lambda(K,N)|$. 
Then from (\ref{UEeq2}) we get 
\begin{align*}
    \frac{E_{d}(\npts^{*}\cap K)}{\tdd(N)}&\le 
    \frac{1}{\tdd(N)}\sum_{i\in \Lambda(K,N)}U_{i,N}(x^{*}_{i,N})\\
    &\le 
\frac{N^{K}}{N}\frac{\gd(S^{d})}{\left(1-r^{d}\gd(S^{d})\right)}+
      \mc{O}_r\left(\frac{1}{\log N}\right).
    \end{align*}
    Letting $N\to \infty$ and then $r\to 0$ in this last inequality, 
    we deduce from (\ref{Sduniformdist}) that inequality (\ref{UEeq1}) holds. 
    
    Now suppose that $A\subset S^{d}$ is almost clopen (with respect 
    to $\mc{H}_{d}$), and let $B:=S^{d}\setminus A$. For any set $K$, we 
    put $K^{N}:=\npts^{*}\cap K$. Then, clearly, 
    \beq
    \label{UEeq3} 
    \frac{E_{d}(\npts^{*})}{\tdd(N)}=\frac{E_{d}(A^{N})}{\tdd(N)} 
    +\frac{E_{d}(B^{N})}{\tdd(N)}+\frac{2}{\tdd(N)}\sum_{\substack{x\in A^{N}\\ %corr 3-1
    y\in B^{N}}}\frac{1}{|x-y|^{d}}. %corr 3-2
    \eeq
    We claim that, as $N\to \infty$, the last term in (\ref{UEeq3}) 
    tends to zero. To see this, let $\epsilon>0$ be given and cover 
    $\partial A$ by an open (relative to $S^{d}$) set 
    $\Omega_{\epsilon}$ such that $\mc{H}_{d}(\Omega_{\epsilon})<\epsilon$ 
    and $\mc{H}_{d}(\partial \Omega_{\epsilon})=0$ ({\it e.g.}, let 
    $\Omega_{\epsilon}$ be a finite union of open balls).  Then, since 
    $\dist (A,B\setminus \Omega_{\epsilon})>0$ and 
    $\dist(B,A\setminus \Omega_{\epsilon})>0$, it follows that, with 
    $\tilde{A}_{\epsilon}:=A\cap \Omega_{\epsilon}$, 
    $\tilde{B}_{\epsilon}:=B\cap \Omega_{\epsilon}$,
    \begin{align} \label{UEeq4}
    \limsup_{N\to \infty}\frac{2}{\tdd(N)}\sum_{\substack{x\in A^{N}\\ 
    y\in B^{N}}}\frac{1}{|x-y|^{d}}&=\limsup_{N\to   %corr 3-3
    \infty}\frac{2}{\tdd(N)}\sum_{\substack{x\in \tilde{A}^{N}\\ 
    y\in \tilde{B}^{N}}}\frac{1}{|x-y|^{d}}\\ \nonumber
    &\le \limsup_{N\to 
    \infty}\frac{1}{\tdd(N)}E_{d}(\bar{\Omega}_{\epsilon}^{N}).
    \end{align}
    Since $\bar{\Omega}_{\epsilon}$ is almost clopen, we get from 
    (\ref{UEeq1}) and (\ref{UEeq4}) that 
    $$
    \limsup_{N\to \infty}\frac{2}{\tdd(N)}\sum_{\substack{x\in A^{N}\\ 
    y\in B^{N}}}\frac{1}{|x-y|^{d}}\le  %corr 3-4
    \gd(S^{d})\frac{\mc{H}_{d}(\bar{\Omega}_{\epsilon})}{\mc{H}_{d}(S^{d})}\le 
    \frac{\epsilon \gd(S^{d})}{\mc{H}_{d}(S^{d})}.
    $$
    As $\epsilon>0$ is arbitrary, we have shown that the last term 
    in (\ref{UEeq3}) goes to zero as $N\to \infty$, as claimed. 
    Consequently, 
    \begin{align}
	\label{UEeq5} \gd(S^{d})=\lim_{N\to 
	\infty}\frac{E_{d}(\npts^{*})}{\tdd(N)}= \lim_{N\to 
	\infty}\left(\frac{E_{d}(A^{N})}{\tdd(N)}+\frac{E_{d}(B^{N})}{\tdd(N)}\right).  %corr 3-5
	\end{align}
	Since $A$ and $B$ are almost clopen and 
	$\mc{H}_{d}(A)+\mc{H}_{d}(B)=\mc{H}_{d}(S^{d})$, it follows from (\ref{UEeq1}) 
	and (\ref{UEeq5}) that 
	$$
	\lim_{N\to\infty}\frac{E_{d}(A^{N})}{\tdd(N)}=\gd(S^{d})\frac{\mc{H}_{d}(A)}{\mc{H}_{d}(S^{d})}
	$$
	and
		$$	\lim_{N\to\infty}\frac{E_{d}(B^{N})}{\tdd(N)}
		=\gd(S^{d})\frac{\mc{H}_{d}(B)}{\mc{H}_{d}(S^{d})}.   %corr 3-6
	$$
	  \eproof
	  
	  \begin{lemma} \label{Ascalable} Suppose $A$ is a compact scalable subset of $S^{d}$. 
	      Then  $\gd(A)$ exists and 
	 $$\gd(A)=\mc{H}_{d}(\mc{B}^{d})/\mc{H}_{d}(A).$$
	  \end{lemma}
	       By saying $A$ is a scalable subset of $S^{d}$, we mean that for 
      every $\epsilon>0$
      there is a bi-Lipschitz mapping  with constant $(1+\epsilon)$ 
      that maps the closure of $A$ into its 
      interior relative to $S^{d}$.  Clearly the measure of the 
      closure of such a 
      set is equal to the measure of its interior, and so any scalable 
      subset of $S^{d}$ is  almost clopen.
  
	  \beginproof
	      Suppose either $C=A$ or $C=B:=S^{d}\setminus A$. Then  (\ref{uniformenergy}) holds.  
	      We first prove that 
	      $\overgd(C)\le\mc{H}_{d}(\mc{B}^{d})/\mc{H}_{d}(C)$.  For $\rho>1$ and 
	      $N\in\N$, let $M[N]:=\lfloor \rho
	      \frac{\mc{H}_{d}(S^{d})}{\mc{H}_{d}(C)}N\rfloor$ where $\lfloor 
	      x\rfloor$ denotes the integer part of $x$.  Let $C^{M[N]}:=
	      \omega_{M[N]}^{*}\cap 
	      C$ and recall that (\ref{Sduniformdist}) states that 
	      $|C^{M[N]}|/M[N]\to \mc{H}_{d}(C)/\mc{H}_{d}(S^{d})$ as $N\to \infty$.
	      
	      Then, for $N$ large enough, we have 
	      $$\rho^{-1}\frac{\mc{H}_{d}(C)}{\mc{H}_{d}(S^{d})}\le 
	      \frac{|C^{M[N]}|}{M[N]}
	      $$
	      from which it follows that there is some $N_{\rho}$ such 
	      that 
	      \beq \label{MNineq}
	      N\le |C^{M[N]}| \qquad (N> N_{\rho}),
	      \eeq
	      and so $\mc{E}_{d}(C,N)\le 
	      E_{d}(\omega_{M[N]}^{*}\cap C)$ for $N> N_{\rho}$.  
	      Thus we have 
	      \begin{align*}
	      \overgd(C)&=\limsup_{N\to 
	      \infty}\frac{\mc{E}_{d}(C,N)}{\tdd(N)}\\ &\le  %corr 4
	      \limsup_{N\to
	      \infty}\frac{\tdd(M[N])}{\tdd(N)}\frac{E_{d}(\omega_{M[N]}^{*}\cap C)}{\tdd(M[N])}\\
	      &\le\rho^{2}\gd(S^{d})\frac{\mc{H}_{d}(S^{d})}{\mc{H}_{d}(C)}, 
	       \end{align*}
	       where (\ref{uniformenergy}) was used to obtain the last inequality. 
	      Since $\rho>1$ is arbitrary, we have $\overgd(C)\le 
	      \mc{H}_{d}(\mc{B}^{d})/\mc{H}_{d}(C)$ for either $C=A$ or $C=B$.

	      Next we show  $\undergd(A)\ge 
	      \mc{H}_{d}(\mc{B}^{d})/\mc{H}_{d}(A)$.  Let
	      $(a_{N})_{N\in\N}$ denote a sequence of 
	      natural numbers such that $\lim_{N\to \infty} \Gd(A,a_{N})=\undergd(A)$. 
	      For $N\in\N$, let $b_{N}=\lceil (\mc{H}_{d}(B)/\mc{H}_{d}(A))a_{N}\rceil$  
	      where $\lceil x \rceil$ denotes the least integer greater than 
	      or equal to $x$ and let $c_{N}=a_{N}+b_{N}$. 
	      Since $A$ is scalable, there is  a bi-Lipschitz mapping $h$ with constant 
	      $(1+\epsilon)$ such that $h(A)\subset 
	      A^{\circ}$.   Then $\delta:=\dist(h(A),B)>0$ and, 
	      as in the proof of Lemma~\ref{unionlemmaupper}, we have 
	       \begin{align*}
     \End(S^{d},c_{N}) &\le \End(h(A)\cup B,c_{N})\le
     \End(h(A),a_{N})+\End(B,b_{N})+2\delta^{-s}a_{N}b_{N}\\
     &\le (1+\epsilon)^{d}\End(A,a_{N})+\End(B,b_{N})+2\delta^{-d}c_{N}^{2}
 \end{align*}  
and thus
$$\Gd(S^{d},c_{N})\le 
(1+\epsilon)^{d}\Gd(A,a_{N})\frac{\tdd(a_{N})}{\tdd(c_{N})}+\Gd(B,b_{N})\frac{\tdd(b_{N})}{\tdd(c_{N})}
+\frac{2\delta^{-d}}{\log(c_{N})}.$$
Letting $N\to \infty$ and then  $\epsilon\to 0$ gives
$$\gd(S^{d})\le 
\undergd(A)\left(\frac{\mc{H}_{d}(A)}{\mc{H}_{d}(S^{d})}\right)^{2}+
\overgd(B)\left(\frac{\mc{H}_{d}(B)}{\mc{H}_{d}(S^{d})}\right)^{2}.$$
Using $\overgd(B)\le 
\mc{H}_{d}(\mc{B}^{d})/\mc{H}_{d}(B)$  and 
$\mc{H}_{d}(S^{d})=\mc{H}_{d}(A)+\mc{H}_{d}(B)$ as well as Theorem~1.2, we get 
$\undergd(A)\ge\mc{H}_{d}(\mc{B}^{d})/\mc{H}_{d}(A)$, which completes the proof of 
	      Lemma~\ref{Ascalable}. 
	  \eproof

	  Now we return to the $s=d$ case of the proof of Theorem~\ref{gU}. 
	  Recall that $\sP:\R^{d}\to S^{d}$ denotes the stereographic 
	projection defined by (\ref{sPdef}).  Let 
	$e_{1}:=(1,0,\ldots,0)\in\R^{d}$ and ${\bf 1}:=(1,1,\ldots,1)\in 
	\R^{d}$.
	For $0<\gamma<1$, let 
	$$U_{\gamma}:=\gamma 
	U^{d}+e_{1}-(\gamma/2){\bf 	1}=
	[1-\gamma/2,1+\gamma/2]\times[-\gamma/2,\gamma/2]\times\cdots\times 
	[-\gamma/2,\gamma/2]$$ and let 
	$A_{\gamma}:=\sP(U_{\gamma})$.  Note that $A_{\gamma}$ is scalable, 
	since the mappings $a\mapsto \sP(r\sP^{-1}(a)+(1-r)e_{1})$, for $0<r<1$,  form a family 
	of bi-Lipschitz mappings with constants approaching 1 as $r\to 1$ 
	that map $A_{\gamma}$ into its relative 
	interior (cf. (\ref{sPdiff})). Thus $\gd(A_{\gamma})$ 
	exists and equals $\mc{H}_{d}(\mc{B}^{d})/\mc{H}_{d}(A_{\gamma})$.   For $x\in U_{\gamma}$ 
	and $0<\gamma<1/d$, we 
	have $1-\gamma\le 
	|x|^{2}\le (1+\gamma/2)^{2}+(d-1)(\gamma/2)^{2}\le 1+2\gamma$.  
	Using (\ref{sPdiff}) it follows that for $\gamma<1/d$, the function $h:=\sP^{-1}$ is %corr 5
	bi-Lipschitz  on $A_{\gamma}$ with constant $(1+\gamma)$ and such that $U_{\gamma}=h(A_{\gamma})$.  Then 
\beq \label{sedEq1}
\overgd(U^{d}) = \gamma^{d}\overgd(U_{\gamma}) = 
\gamma^{d}\overgd(h(A_{\gamma}))\le \gamma^{d}(1+\gamma)^{d}\gd(A_{\gamma})
\eeq
	and, similarly,
\beq \label{sedEq2}
	\undergd(U^{d}) = \gamma^{d}\undergd(U_{\gamma}) \ge 
 \gamma^{d}(1+\gamma)^{-d}\gd(A_{\gamma}).
	\eeq
Since  $h=\sP^{-1}$ is bi-Lipschitz on  $A_{\gamma}$ with constant $(1+\gamma)$, it follows 
	that $\lim_{\gamma\to 0^{+}}\gamma^{-d}\mc{H}_{d}(A_{\gamma})=\mc{H}_{d}(U^{d})=1$ 
	and so 
$$\gamma^{d}\gd(A_{\gamma})=\gamma^{d}\mc{H}_{d}(\mc{B}^{d})/\mc{H}_{d}(A_{\gamma})\to \mc{H}_{d}(\mc{B}^{d})$$
as $\gamma\to 0$.  Taking $\gamma\to 0$ in (\ref{sedEq1}) and 
(\ref{sedEq2}) we then have
$$
\mc{H}_{d}(\mc{B}^{d})\le \undergd(U^{d})\le \overgd(U^{d})\le \mc{H}_{d}(\mc{B}^{d})
$$
which completes the proof of Theorem~\ref{gU}. 	  
     
   \section{Almost clopen sets in $\R^{d}$}
A  Lebesgue measurable set $A\subset \R^{d}$ is said to be 
{\bf almost clopen (with respect to $d$-dimensional Lebesgue measure)} if
$\mc{H}_{d}(\partial A)=0$ where $\partial A$
denotes the boundary of $A$.  

\begin{theorem} \label{almostclopenThm}
Suppose $A$ is a bounded almost clopen set in $\R^{d}$. Then 
$\g(A)$ exists for $s\ge d$ and 
\beq\label{gm}
\g(A)=\g(U^{d})\mc{H}_{d}(A)^{-s/d}.
\eeq
\end{theorem}

{\noindent \bf Remark.} In particular, $\g(A)=\infty$ if $\mc{H}_{d}(A)=0$.

\beginproof
    First, if $A=\gamma U^{d}$ then 
    $\g(A)=\gamma^{-s}\g(U^{d})=\mc{H}_{d}(A)^{-s/d}\g(U^{d})$ showing that $A$ satisfies (\ref{gm}).
    Applying Corollary~\ref{AcupBCorollary} inductively, it then follows that (\ref{gm}) holds if $A$ is the 
    union of a finite collection of cubes with disjoint interiors. 
    
Next,  for $n\in\N$, let $\mc{Q}_{n}$ denote the cubes $q$ in $\R^{d}$ with
vertices in the lattice $\Z^{d}/n$,  let 
$\underline{A}_{n}$ denote the union of the cubes in 
$\mc{Q}_{n}$ that are also contained in 
$A$ and let $\overline{A}_{n}$ denote the union of the cubes $\mc{Q}_{n}$ that 
meet the closure of $A$.  

Suppose $\epsilon>0$, then there is an open set $V$ containing 
$\partial A$ with $\mc{H}_{d}(V)<\epsilon$. If $q\in\mc{Q}_{n}$ is a 
subset of $\overline{A}_{n}\cap\underline{A}_{n}^{c}$, the complement
of $\underline{A}_{n}$ in $\overline{A}_{n}$, then $q$ meets $\partial A$.
Since $\partial A$ is compact and $V^{c}$ is closed, the distance 
$\text{dist }(\partial A,V^{c})>0$.

Let $n^{*}$ be large enough so that $\text{diam }q<\text{dist }(\partial 
A,V^{c})$ for $q\in \mc{Q}_{n^{*}}$.
If $q\in \mc{Q}_{n^{*}}$ meets $\partial A$ then $q\subset V$ and so we 
have
$\overline{A}_{n}\cap\underline{A}_{n}^{c}\subset V$ for $n>n^{*}$. 
Hence,
$$\mc{H}_{d}(\underline{A}_{n})\le \mc{H}_{d}(\overline{A}_{n}) \le 
\mc{H}_{d}(\underline{A}_{n})+\epsilon  \qquad (n>n^{*})
$$
showing that $$\lim_{n\to \infty}\mc{H}_{d}(\underline{A}_{n})=
\lim_{n\to \infty}\mc{H}_{d}(\overline{A}_{n}) =\mc{H}_{d}(A).$$
Since $\g(\overline{A}_{n})\le \underg(A)\le \overg(A) \le \g(\underline{A}_{n})$
and (\ref{gm}) holds for $\overline{A}_{n}$ and $\underline{A}_{n}$ it 
follows that (\ref{gm}) holds for $A$. 
\eproof

We say that a sequence $\npts \subset A$, $N\in \mc{N}$, of sets of points 
in $A$ is {\bf asymptotically $s$-energy minimizing on $A$} for $s\ge 
d$ if $$\lim_{N\to \infty} 
\frac{\E(\npts)}{\tsd(N)}=\underg(A) \qquad (N\in\mc{N}).$$

\begin{corollary}
  Suppose $A$ is a bounded, almost clopen set in $\R^{d}$, 
  $\mc{H}_{d}(A)>0$, 
  $B$ is an almost clopen subset of $A$ and that
$\npts \subset A$, $N\in \mc{N}$, is 
asymptotically $s$-energy minimizing on $A$.  Then we have
\beq\label{unifdist2}
\lim_{\substack{N\to \infty\\ N\in \mc{N}}}\frac{|\npts\cap B|}{N}=\mc{H}_{d}(B)/\mc{H}_{d}(A) .
\qquad (N\in\mc{N})
\eeq
\end{corollary}

\beginproof
    Note that $B':=A\setminus B$ is almost clopen (since  $\partial 
    B'\subset \partial A\cup \partial B$).  Applying 
    Theorem~\ref{almostclopenThm} to $A$, $B$ and $B'$ gives  
  \begin{align*}
  \g(A)&=\g(U^{d})\mc{H}_{d}(A)^{-s/d}\\ &=\g(U^{d})(\mc{H}_{d}(B)+\mc{H}_{d}(B'))^{-s/d}=
  \left(\g(B)^{-d/s}+\g(B')^{-d/s}\right)^{-s/d}.
  \end{align*}
    Then Lemma~\ref{unionlemmalower}  implies (\ref{unifdist2}).
  \eproof

  \section{Separation}
%\beginproof
\noindent {\it  Proof of Theorem~\ref{main4}.}
    For convenience we denote $x_{i,N}^{*}$ by $x_{i}$.  For 
    $i=1,\ldots,N$, let 
    \beq \label{Uidef}
    U_{i}(x):=\sum_{j\neq i}\frac{1}{|x-x_{j}|^{s}}.
    \eeq
    Then $U_{i}(x_{i})\le U_{i}(x)$ for all $x\in A$.  Let 
    $0<\delta < 1$ and set  $$r_{0}:=(\delta \mc{H}_{d}(A)/(N \mc{H}_{d}(B(0,1))))^{1/d}$$ 
    and 
    $$
    D_{j}:=B(x_{j},r_{0}),\quad 
    \mc{D}_{i}:=A\setminus \bigcup_{j\neq i}D_{j}.  
    $$
    Then 
    \beq \label{sepbnd1}
    \mc{H}_{d}(\mc{D}_{i})\ge 
    \mc{H}_{d}(A)-Nr_{0}^{d}\mc{H}_{d}(B(0,1))=\mc{H}_{d}(A)(1-\delta)>0
    \eeq
    and we have 
    \begin{align}
	U_{i}(x_{i})&\le 	
	\frac{1}{\mc{H}_{d}(\mc{D}_{i})}\int_{\mc{D}_{i}}U_{i}(x)\,d \mc{H}_{d}(x)\\ \nonumber
	&=\frac{1}{\mc{H}_{d}(\mc{D}_{i})}\sum_{j\neq 
	i}\int_{\mc{D}_{i}}\frac{1}{|x-x_{j}|^{s}}\,d \mc{H}_{d}(x)\\ 	\nonumber
	&\le \frac{1}{\mc{H}_{d}(\mc{D}_{i})}\sum_{j\neq i}\int_{A\setminus 
	D_{j}}\frac{1}{|x-x_{j}|^{s}}\,d \mc{H}_{d}(x).
    \end{align}
    Let $R>\diam A$.  It is easy to verify that for $0<r<1$ and $y\in 
    A$ 
  \begin{align}
  \label{sepbnd2}\int_{A\setminus 
	B(y,r)}\frac{1}{|x-y|^{s}}\,d \mc{H}_{d}(x)&\le \int_{B(0,R)\setminus 
	B(0,r)}\frac{1}{|u|^{s}}\, d \mc{H}_{d}(u)\\
	&\le \nonumber
	\begin{cases}
	    c_{s}/r^{s-d} & \text{for $s>d$,}\\
	    c_{d}\log(R/r) & \text{for $s=d$,}
	    \end{cases}
	\end{align}
	where the positive constants $c_{s}$, $c_{d}$ are independent of $y$ 
	and $r$.  Using the estimates (\ref{sepbnd1}) and (\ref{sepbnd2}) we 
	get for $s>d$,
	\beq \label{sepbnd3}
	U_{i}(x_{i})\le 
	\frac{(N-1)c_{s}\left(\delta \mc{H}_{d}(A)/N \mc{H}_{d}(B(0,1))\right)^{1-s/d}}{(1-\delta)\mc{H}_{d}(A)}\le 
	k_{s}N^{s/d}\eeq
	and for $s=d$, 
	\beq\label{sepbnd4}
	U_{i}(x_{i})\le k_{d}N\log N,
	\eeq
	where the constants $k_{s}$, $k_{d}$ are independent of $N$ and 
	$i$.  Finally, since for each $i=1,\ldots, N$,
	we have $|x_{i}-x_{j}|^{-s}\le U_{i}(x_{i})$ for $i\neq j$, 
	inequality (\ref{sepcond}) follows from (\ref{sepbnd3}) and (\ref{sepbnd4}).
    \eproof

    \begin{lemma}\label{sepsed} For the closed  unit ball $\mc{B}^{d}:={\bar 
    B}(0,1)\subset\R^{d}$,  
	there is a $d$-energy asymptotically optimal sequence $(\npts)_{N\in\N}$ of $N$-point 
	configurations $\npts=\{x_{1,N},\ldots,x_{N,N}\}$ for $\mc{B}^{d}$ such that  for $N\ge 2$
 $$\min_{i\neq j}|x_{i,N}-x_{j,N}|\ge (2+\sqrt{d})^{-1}N^{-1/d}.$$  
    \end{lemma}
    \beginproof
	By Theorems~\ref{gU} and~\ref{almostclopenThm}, we have that 
	$\gd(\mc{B}^{d})=1$.
For a positive integer $m$ let
$\Omega^{m}:=(\frac{1}{m}\Z)^{d}\cap \mc{B}^{d}$ and for ${\bf j}\in (\frac{1}{m}\Z)^{d}$ let 
$U_{m,{\bf j}}:=\frac{1}{m}[-1/2,1/2]^{d}+{\bf j}$ denote the $d$ dimensional cube 
of side length $1/m$ (with sides parallel to the coordinate axes) and 
center ${\bf j}$.  Since 
$$
B(0,(1-\sqrt{d}/m))\subset \bigcup_{{\bf j}\in \Omega^{m}}U_{m,{\bf j}}\subset 
B(0,(1+\sqrt{d}/m)),
$$
we have 
\beq \label{cardOmega}
(m-\sqrt{d})^{d}\mc{H}_{d}(\mc{B}^{d})\le |\Omega^{m}|\le (m+\sqrt{d})^{d}\mc{H}_{d}(\mc{B}^{d}).
\eeq
Fix $k>\sqrt{d}$.
If $x\in U_{m,{\bf j}}$ and $|{\bf j}|\ge k/m$, then $ |{\bf j}|\ge |x|-\sqrt{d}/(2m)>0$ and so 
\begin{align*}
\sum_{\substack{{\bf j}\in(\frac{1}{m}\Z)^{d}\\ 0<|{\bf j}|\le 2}}\frac{1}{|{\bf j}|^{d}}
&\le \sum_{\substack{{\bf j}\in(\frac{1}{m}\Z)^{d}\\ 0<|{\bf j}|< 
k/m}}\frac{1}{|{\bf j}|^{d}} + 
m^{d}\sum_{\substack{{\bf j}\in(\frac{1}{m}\Z)^{d}\\ k/m\le |{\bf 
j}|\le 
2}}\frac{1}{|{\bf j}|^{d}}\frac{1}{m^{d}}\\
& \le \sum_{\substack{{\bf j}\in(\frac{1}{m}\Z)^{d}\\ 0<|{\bf j}|< k/m}}\frac{1}{|{\bf j}|^{d}}
+m^{d}\int_{k/m<|x|< 2}\frac{1}{(|x|-\sqrt{d}/(2m))^{d}}\, d \mc{H}_{d}(x)\\
&\le 2^{d}k^{d}m^{d}+
m^{d}\int_{(k-\sqrt{d}/2)/m}^{2}\frac{(r+\sqrt{d}/(2m))^{d-1}}{r^{d}}\mc{H}_{d-1}(S^{d-1})dr\\
&\le 2^{d}k^{d}m^{d}+
m^{d}(1+\sqrt{d}/k)^{d-1}\mc{H}_{d-1}(S^{d-1})\int_{(k-\sqrt{d}/2)/m}^{2}\frac{1}{r}dr \\
&= 2^{d}k^{d}m^{d} +m^{d}\mc{H}_{d-1}(S^{d-1})(1+\sqrt{d}/k)^{d-1}\log \left(\frac{2m}{k-\sqrt{d}/2}\right). 
\end{align*}
Hence using (\ref{cardOmega}) and the preceding estimate we obtain
\begin{align}
&\Ed(\Omega^{m})=\sum_{i\in\Omega^{m}}\sum_{\substack{{\bf j}\in\Omega^{m} \nonumber \\{\bf j}\neq 
i}}\frac{1}{|{\bf j}-{\bf i}|^{d}}  \le
|\Omega^{m}|\sum_{\substack{{\bf j}\in(\frac{1}{m}\Z)^{d}\\ 0<|{\bf 
j}|\le 
2}}\frac{1}{|{\bf j}|^{d}}\\ \nonumber &\le
m^{d}(m+\sqrt{d})^{d}\mc{H}_{d}(\mc{B}^{d})\left(2^{d}k^{d} +\mc{H}_{d-1}(S^{d-1})(1+\sqrt{d}/k)^{d-1}\log \left(\frac{2m}{k-\sqrt{d}/2}\right)\right).
\end{align}

Suppose $N\ge 2$.  Now choose $m=\lceil 
(N/\mc{H}_{d}(\mc{B}^{d}))^{1/d}+\sqrt{d}\rceil$. 
Then using (\ref{cardOmega}) we get
\beq \label{sedsepNm}
(m-\sqrt{d}-1)^{d}\mc{H}_{d}(\mc{B}^{d})\le N\le (m-\sqrt{d})^{d}\mc{H}_{d}(\mc{B}^{d})\le |\Omega^{m}|.
\eeq
Hence we may let $\npts$ consist of $N$ distinct points from $\Omega^{m}$.  
Then
\begin{align*}
&\frac{\Ed(\npts)}{N^{2}\log N}\le \frac{\Ed(\Omega^{m})}{N^{2}\log 
N}\\ & \le 
\left(\frac{m^{d}(m+\sqrt{d})^{d}}{(m-\sqrt{d}-1)^{2d}}\right)
\frac{2^{d}k^{d} +\mc{H}_{d-1}(S^{d-1})(1+\sqrt{d}/k)^{d-1}
\log \left(\frac{2m}{k-\sqrt{d}/2}\right)}{\mc{H}_{d}(\mc{B}^{d}) \log((m-\sqrt{d}-1)^{d}\mc{H}_{d}(\mc{B}^{d}))}.
\end{align*}
On taking $N\to \infty$ (and thus $m\to \infty$) we get
$$
\limsup_{N\to \infty}\frac{\Ed(\npts)}{N^{2}\log N}\le 
\frac{\mc{H}_{d-1}(S^{d-1})}{d\, \mc{H}_{d}(\mc{B}^{d})}(1+\sqrt{d}/k)^{d-1}=(1+\sqrt{d}/k)^{d-1}
$$
for any $k\ge \sqrt{d}$ (here we recall $\mc{H}_{d-1}(S^{d-1})=d\, 
\mc{H}_{d}(\mc{B}^{d})$).  Letting $k\to \infty$ then shows
that $(\npts)_{N\in\N}$ is  
$d$-energy asymptotically  optimal for $\mc{B}^{d}$.  Using  $\mc{H}_{d}(\mc{B}^{d})\ge 1$ and the definition of $m$,  
we have $m\le N^{1/d}(2+\sqrt{d})$ and thus 
$$\min_{x\neq y\in \npts}|x-y|=1/m\ge (2+\sqrt{d})^{-1}N^{-1/d} $$ which 
completes the proof. 
\eproof

\section{Compact sets}
%\beginproof
\noindent {\it  Proof of Theorem~\ref{main1}.}   
Let $\epsilon>0$ and $G$ be an almost clopen set (since 
$A$ is compact,  $G$ could 
be chosen to be the union of a finite collection of open balls) such that $G\supset A$	and 
$\mc{H}_{d}(G\setminus A)<\epsilon$.   Then, from 
Theorem~\ref{almostclopenThm}, 
\beq \label{AGineq}
\underg(A)\ge \g(G)=\g(U^{d})\mc{H}_{d}(G)^{-s/d}\ge 
\g(U^{d})(\mc{H}_{d}(A)+\epsilon)^{-s/d}.
\eeq
If $\mc{H}_{d}(A)=0$ then (\ref{AGineq}) shows $\underg(A)=\overg(A)=\infty$; if 
 $\mc{H}_{d}(A)>0$, 
then since (\ref{AGineq}) holds for arbitrary $\epsilon>0$,  
we get
\beq
\label{compact0}
\underg(A)\ge \g(U^{d})\mc{H}_{d}(A)^{-s/d}.
\eeq

\bigskip

We next show $\overg(A)\le \g(U^{d})\mc{H}_{d}(A)^{-s/d}$. The case 
$\mc{H}_{d}(A)=0$ was already considered above and so we assume  $\mc{H}_{d}(A)>0$.  Let 
$$A^{*}:=\{x\in A\mid \limsup_{r\to 0^{+}}\frac{\mc{H}_{d}(\bar B(x,r)\cap 
A)}{\mc{H}_{d}(\bar B(x,r))}=1\}.$$
The Lebesgue Density Theorem  (e.g., see \cite{Morgan}) states that
$\mc{H}_{d}(A\setminus A^{*})=0$.
For $0<\epsilon<1$, let 
\beq \label{Ceps}
C_{\epsilon}:=\{{\bar B}(x,r)\mid x\in A^{*},\,0<r<1,\,  \frac{\mc{H}_{d}({\bar B}(x,r)\cap 
A)}{\mc{H}_{d}({\bar B}(x,r))}>1-\epsilon \}.
\eeq

By the Besicovitch Covering Theorem (cf. \cite{Morgan}), there is a 
countable collection of pairwise disjoint closed  balls 
$\{B_{i}:={\bar B}(x_{i},r_{i})\}\subset C_{\epsilon}$ that covers almost all of $A^{*}$
and hence almost all of $A$.  
Choose $n$ large enough so that 
\beq \label{cupBi}
\mc{H}_{d}\left(\bigcup_{i=1}^{n}A\cap 
B_{i}\right)=\sum_{i=1}^{n}\mc{H}_{d}(A\cap B_{i})\ge 
(1-\epsilon)\mc{H}_{d}(A).
\eeq 

Let $i\in\{1,\ldots,n\}$ be fixed and
let $\npts$ denote an asymptotically minimal  sequence of 
configurations for $B_{i}$ such that $$\delta_{N}:=\min_{x,y\in \npts, \, x\neq y}|x-y|\ge 
r_i(CN)^{-1/d}$$ for some positive constant $C$ independent of $i$. (Recall 
Theorem~\ref{main4} states  that, in the case  
$s>d$,  any minimal 
sequence for $B_{i}$ must satisfy such a separation condition  while 
Lemma~\ref{sepsed} implies the 
existence of such a sequence 
in the case $s=d$.)

For $0<\nu<1/2$, let $r:=\nu\delta_{N}$ and set $$\npts^{\nu}:=\{x\in \npts\mid 
\dist(x,A\cap B_{i})\le r\}.$$  
Then 
$$B(x,r)\cap B(y,r)=\emptyset\qquad \text{ for }\,  x,y\in \npts, \; x\neq y,$$
and 
$$B(x,r)\cap A=\emptyset \qquad \text{ for }\,  x\in \npts\setminus \npts^{\nu}.$$ 

Since, for any fixed constant less than 1/2, say 
1/4, at least this fraction of every $B(x,r)$, $x \in B_i$, is contained in $B_i $
for $N$ sufficiently large (and hence $r$ sufficiently small), we have
$$\mc{H}_{d}(B_{i}\cap A^{c})\ge \mc{H}_{d}(\bigcup_{x\in \npts\setminus 
\npts^{\nu} }B_{i}\cap B(x,r))\ge (1/4)|\npts\setminus 
\npts^{\nu}|\, \mc{H}_{d}(B(0,1))r^{d},$$ which implies
$$
|\npts\setminus 
\npts^{\nu}|\le 4\mc{H}_{d}(B_{i}\cap 
A^{c})\mc{H}_{d}(B(0,1))^{-1}(\nu\delta_{N})^{-d}\le
4C\frac{\epsilon}{\nu^{d}}N,
$$
where we have used (cf. (\ref{Ceps})) $\mc{H}_{d}(B_{i}\cap 
A^{c})\le \epsilon \mc{H}_{d}(B_{i})= \epsilon r_i^d  \mc{H}_{d}(B(0,1))$.  Thus
\beq \label{compacteq1}
|\npts^{\nu}|=N-|\npts\setminus 
\npts^{\nu}|\ge N(1-4C\epsilon\nu^{-d}).
\eeq

If $x\in \npts^{\nu}$, then there exists $y\in A\cap B_{i}$ such that 
$|x-y|\le r$. For each $x\in \npts^{\nu}$, let 
$\phi_{N,\nu}(x)$ be one such $y$, and let
$$\lambda_{N,\nu}:=\{\phi_{N,\nu}(x)\mid x\in \npts^{\nu}\}.$$  
Note that for $x,y\in \npts^{\nu}$ we have 
\begin{align}
|\phi_{N,\nu}(x)-\phi_{N,\nu}(y)|&\ge 
|x-y|-|\phi_{N,\nu}(x)-x|-|\phi_{N,\nu}(y)-y|\\
&\ge (1-2\nu)|x-y|,\nonumber
\end{align}
and so 
\beq
\E(\lambda_{N,\nu})\le (1-2\nu)^{-s}\E(\npts).
\eeq

Let $M:=\lceil N/ (1-4C\epsilon \nu^{-d})\rceil$.  Then
$$|\lambda_{M,\nu}|\ge (1-4C\epsilon \nu^{-d})M\ge N,$$
and so we have
\begin{align}
\frac{\En(A\cap B_{i},N)}{\tsd(N)}&\le 
\left(\frac{\E(\lambda_{M,\nu})}{\tsd(M)}\right)\left(\frac{\tsd(M)}{\tsd(N)} \right)\\
&\le  \nonumber
\frac{1}{(1-2\nu)^{s}}\left(\frac{\tsd(M)}{\tsd(N)}\right)
\left(\frac{\E(\omega_{M})}{\tsd(M)}\right).
\end{align}
From the definition of $M$ it follows (even in the case $s=d$) that 
$$\lim_{N\to \infty}\tsd(M)/\tsd(N)= \left(\frac{1}{1-4C\epsilon 
\nu^{-d}}\right)^{1+s/d}
$$
and hence
\beq \label{compact2}
\overg(A\cap B_{i})\le \frac{1}{(1-2\nu)^{s}}\left(\frac{1}{1-4C\epsilon 
\nu^{-d}}\right)^{1+s/d}\g(B_{i})
\eeq
for any $(4C\epsilon)^{1/d}<\nu<1/2$.

Now Theorem~\ref{almostclopenThm} implies 
$\g(B_{i})=\g(U^{d})\mc{H}_{d}(B_{i})^{-s/d}$ and so
using Lemma~\ref{unionlemmaupper} and inequalities  (\ref{compact2}) 
and (\ref{cupBi}) we obtain
\begin{align} \label{compact3}
\overg(A)&\le \overg\left(\bigcup_{i=1}^{n}A\cap B_{i}\right)\\ &\le \nonumber
\left(\sum_{i=1}^{n}\overg(A\cap B_{i})^{-d/s}\right)^{-s/d}\\ &\le  \nonumber
\frac{1}{(1-2\nu)^{s}}\left(\frac{1}{1-4C\epsilon 
\nu^{-d}}\right)^{1+s/d}\g(U^{d})\left(\sum_{i=1}^{n}\mc{H}_{d}(B_{i})\right)^{-s/d}\\ \nonumber 
&\le\frac{1}{(1-2\nu)^{s}}\left(\frac{1}{1-4C\epsilon 
\nu^{-d}}\right)^{1+s/d} \g(U^{d}) (1-\epsilon)^{-s/d} \mc{H}_{d}(A)^{-s/d} 
\end{align}
for any $0<\epsilon<1$ and any 
$(4C\epsilon)^{1/d}<\nu<1/2$.  By first taking $\epsilon\to 0$ and then $\nu\to 
0$ we have
\beq \label{compact4}
\overg(A)\le \g(U^{d})\mc{H}_{d}(A)^{-s/d}
\eeq
which combined with (\ref{compact0}) completes the proof.  
\eproof

%\beginproof
\noindent {\it  Proof of Theorem~\ref{main2}.}
    Let $B\subset A$ be a measurable set such that 
    $\mc{H}_{d}(\partial_{r}B)=0$, where $\partial_{r}B:=\partial 
    B\cap \overline{(A\setminus B)}$ is the relative boundary of $B$. 
    Then $A=A_{1}\cup A_{2}$, where $A_{1}:=B\cup \partial_{r}B$,
    and $A_{2}:=(A\setminus B)\cup \partial_{r}(A\setminus B)$
    are compact sets and $\mc{H}_{d}(\partial_{r}(A\setminus B))=0$. 
    Since $\mc{H}_{d}(A)=\mc{H}_{d}(A_{1})+\mc{H}_{d}(A_{2})$ we can 
    apply Theorem~\ref{main1} and Lemma~\ref{unionlemmalower} to 
    deduce that
    \beq \label{main2proof1}
    \frac{|\npts \cap A_{1}|}{N} \longrightarrow 
    \frac{\mc{H}_{d}(B)}{\mc{H}_{d}(A)} \quad \text{ as $N\to \infty$.}
    \eeq
    On writing $A=\partial_{r}B \cup \overline{A\setminus 
    \partial_{r}B}$ we similarly have
    $$
     \frac{|\npts \cap \partial_{r}B|}{N} \longrightarrow 
    0 \quad \text{ as $N\to \infty$}
    $$
    which together with (\ref{main2proof1}) gives (\ref{wconv2}) and 
    thus (\ref{wconv}).
\eproof

\section{$d$-Rectifiable manifolds in $\R^{d'}$}
In Section~2 we defined the notion of a $d$-rectifiable manifold. More 
generally, 
      a set $A\subset \R^{d'}$ is said to be a {\bf $d$-dimensional rectifiable 
      set} if $A$ is $\mc{H}_{d}$ measurable, $\mc{H}_{d}(A)<\infty$,
      and $\mc{H}_{d}$-almost all of $A$ is contained in the countable 
      union of Lipschitz images of bounded subsets of $\R^{d}$
      (see \cite{Fed}, \cite{Mat}, and \cite{Morgan}).
      Clearly, any $d$-rectifiable manifold is a $d$-dimensional
      rectifiable set. 

% A set $A\subset \R^{d'}$ is said to be a {\bf $d$-rectifiable 
% set} if $A=\phi(B)$ for some Lipschitz mapping $\phi:\R^{d}\to\R^{d'}$ and some 
% bounded set $B\subset \R^{d}$ (cf. \cite[3.2.14]{Fed}).  
% Clearly, any $d$-rectifiable manifold is a $d$-rectifiable set. 

We shall need the following result of Federer  concerning 
 $d$-dimensional rectifiable sets:  
  \begin{lemma}[{\cite[3.2.18]{Fed},  \cite[3.11]{Morgan}}] 
  \label{drectlemma}
    Suppose $A\subset \R^{d'}$ is a $d$-dimensional rectifiable set and 
    $\epsilon>0$. Then there exists a countable collection 
    $\{K_{i}\mid i=1,2,\ldots\}$ of compact subsets of $\R^{d}$
    and  bi-Lipschitz mappings $\psi_{i}:K_{i}\to \R^{d'}$, $i=1,2,\ldots$,
    with constant $(1+\epsilon)$  such that 
    $\psi_{1}(K_{1}),\psi_{2}(K_{2}), \psi_{3}(K_{3})\ldots$ are pairwise disjoint subsets of $A$
    that cover $\mc{H}_{d}$-almost all of $A$.  
    \end{lemma}

\begin{proposition}\label{prop1}
Suppose $A\subset\R^{d'}$ is a compact, $d$-dimensional rectifiable set and 
$s\ge d$.  
Then 
\beq \label{propover0}
\overg(A)\le\g(U^{d})\mc{H}_{d}(A)^{-s/d}.
\eeq
\end{proposition}
\beginproof
    If $\mc{H}_{d}(A)=0$, then the right hand side of (\ref{propover0}) 
    is understood to be $\infty$ and  (\ref{propover0}) holds 
    trivially.  
    Now suppose $0<\epsilon<\mc{H}_{d}(A)$. Let $K_{1},K_{2},\ldots$ and 
    $\psi_{1},\psi_{2},\ldots$ be as in Lemma~\ref{drectlemma}.  
    Let $n\in\N$ be large enough so that
 \beq
 \label{propover1}
 \sum_{i=1}^{n}\mc{H}_{d}(\psi_{i}(K_{i}))\ge 
 \mc{H}_{d}(A)-\epsilon.
\eeq 

Since $\psi_{1}(K_{1}),\ldots,\psi_{n}(K_{n})$ are disjoint compact subsets 
 of $A$, we may use Lemma~\ref{unionlemmaupper}, Theorem~\ref{main1},
 (\ref{propover1}), and the fact that $\psi_{i}$ is bi-Lipschitz with 
 constant $(1+\epsilon)$  to get
 \begin{align}
   \overg(A)&\le \label{propover2}
   \overg\left(\bigcup_{i=1}^{n}\psi_{i}(K_{i})\right)  \\ \nonumber
   &\le \left(\sum_{i=1}^{n}\overg(\psi_{i}(K_{i}))^{-d/s}\right)^{-s/d}   
   \\ 
   &\le   
    \g(U^{d})(1+\epsilon)^{2s}\left(\sum_{i=1}^{n}\mc{H}_{d}(\psi_{i}(K_{i}))\right)^{-s/d}\nonumber \\
   &\le \g(U^{d})(1+\epsilon)^{2s} (\mc{H}_{d}(A)-\epsilon)^{-s/d}.\nonumber
  \end{align}
Since $\epsilon$ is arbitrary, (\ref{propover2}) shows $\overg(A)\le 
\g(U^{d})\mc{H}_{d}(A)^{-s/d}$.
\eproof
    
  \begin{proposition}\label{prop2}
Suppose $s\ge d$ and that $A\subset\R^{d'}$ is as in Theorem~\ref{main3} with the property that for each $\epsilon>0$ there is some $\delta>0$ such that
$\underg(B)\ge 1/\epsilon$ whenever $B$ is a compact subset of $A$ with $\mc{H}_{d}(B)<\delta$.
Then $\underg(A)\ge\g(U^{d})\mc{H}_{d}(A)^{-s/d}$.
\end{proposition}
  
\beginproof
Suppose $\epsilon>0$.   Again let $(K_{i}, \psi_{i})$, $i=1,2,\ldots$, be as in 
 Lemma~\ref{drectlemma}.  
Let $\delta>0$ be    such 
 that $\underg(B)\ge (\epsilon  )^{-s/d}$ whenever $B$ is a compact subset 
 of $A$ with $\mc{H}_{d}(B)<\delta$.
  Let $n$ be large 
 enough so that 
 \beq\label{propunder1}
 \sum_{i=1}^{n}\mc{H}_{d}(\psi_{i}(K_{i}))\ge 
 \mc{H}_{d}(A)-\delta.
\eeq
      Since  $A$ is a Borel set and $\mc{H}_{d}(A)<\infty$,  then $\mc{H}_{d}|_{A}$ is 
    a Radon measure on $\R^{d'}$ (cf. \cite[1.11]{Mat}). If $K\subset 
    A$   is compact and  $\epsilon>0$, then there is some relatively open set $G\subset A$ 
    such that $K\subset G$ and such that $\mc{H}_{d}(G)\le 
    \mc{H}_{d}(K)+\epsilon$. Furthermore, we may choose $G$ to be 
    $\mc{H}_{d}$-almost clopen relative to $A$. Indeed, if $G$ is not almost clopen then we 
    can construct an almost clopen set $\mc{G}$ with the same 
    properties as $G$ in the following way. Let 
    $C(x,r)=\{y\in\R^{d'}\mid |y-x|=r\}$. 
    Since $\mc{H}_{d}(A)<\infty$,  the set $\{r>0\mid 
    \mc{H}_{d}(C(x,r)\cap A)>0\}$ is at most countable. Since 
    $K\subset A$ is 
    compact, there is a relatively open cover of $K$ of the form
    $\{B(x_{i},r_{i})\cap A\mid i=1,\ldots,m\}$ where 
    $B(x_{i},r_{i})\cap A\subset G$ and $\mc{H}_{d}( 
    C(x_{i},r_{i})\cap A)=0$. Let 
    $\mc{G}=\bigcup_{i=1}^{m}B(x_{i},r_{i})\cap A$, then $K\subset 
    \mc{G}\subset G$, $\mc{G}$ is a 
    relatively open subset of $A$, and $\mc{H}_{d}(\bar {\mc{G}}) 
    =\mc{H}_{d}(\mc{G})\le \mc{H}_{d}(K)+\epsilon$.
  
 Using arguments\footnote{For further details, see the online addendum {\tt arXiv:math-ph/0412053}.} similar to those in the proof of Theorem~2.1 for  $s>d$ (or the assumption that $A$ is a subset of $d$-dimensional $C^1$ manifold in the case $s=d$), we can find for $i=1,\ldots, n$   a relatively 
    open subset $G_{i}$ of $A$   such that $\psi_{i}(K_{i})\subset G_{i}$ and
  \begin{align}
   \label{propunder2}
    \underg(\bar G_{i})&\ge 
        \left(\underg(\psi_{i}(K_{i}))^{-d/s}+\epsilon/2^{i}\right)^{-s/d}. 
  \end{align}

    Let $G_{0}:=A\setminus\bigcup_{i=1}^{n}\bar G_{i}$.  Then $\bar 
    G_{0}\subset A\setminus\bigcup_{i=1}^{n}\psi_{i}(K_{i})$ and thus, using 
    (\ref{propunder1}), we obtain
    $\mc{H}_{d}(\bar G_{0})\le \delta_{0}$ and hence
  \begin{equation}
  \underg(G_{0})\ge \epsilon^{-s/d}. 
  \end{equation}
  
           Since $\psi_{i}$   is bi-Lipschitz on $K_{i}$ with constant 
   $(1+\epsilon)$ we have, using Theorem~\ref{main1}, 
    \begin{align}  \label{propunder3}
  \underg(\psi_{i}(K_{i}))&\ge 
  (1+\epsilon)^{-s}\g(K_{i})\\ 
  &=(1+\epsilon)^{-s}\g(U^{d})\mc{H}_{d}(K_{i})^{-s/d}
  \nonumber \\
  &\ge 
  (1+\epsilon)^{-2s}\g(U^{d})\mc{H}_{d}(\psi_{i}(K_{i}))^{-s/d}.\nonumber
   \end{align}

  Since $A\subset \bigcup_{i=0}^{n}\bar G_{i}$, we again use 
  Lemma~\ref{unionlemmalower} together with (82)--(85) to obtain 
  \begin{align}
     \underg(A)&\ge \label{propunder4}
     \left(\sum_{i=0}^{n}\underg(\bar 
     G_{i})^{-d/s}\right)^{-s/d} \\
     &\ge \left(\sum_{i=0}^{n}\epsilon/2^{i}+\sum_{i=1}^{n}\underg(
    \psi_{i}(K_{i}))^{-d/s}\right)^{-s/d}\nonumber\\
     &\ge    \left(2\epsilon+(1+\epsilon)^{2d}\g(U^{d})^{-d/s}\sum_{i=1}^{n}\mc{H}_{d}(\psi_{i}(K_{i}))
     \right)^{-s/d}      \nonumber \\
     &\ge  
     \left(2\epsilon+(1+\epsilon)^{2d}\g(U^{d})^{-d/s}\mc{H}_{d}(A)
     \right)^{-s/d} .     \nonumber
     \end{align}
  Taking  $\epsilon\to 0$ in (\ref{propunder4}) then completes the proof. \eproof

%\beginproof
\noindent {\it Proof of Theorem~\ref{main3}.}
    Suppose $A\subset \R^{d'}$ is a $d$-rectifiable manifold. Since %corr 6
 any $d$-rectifiable manifold is a $d$-dimensional rectifiable set, 
 Proposition~\ref{prop1} implies
 $\overg(A)\le\g(U^{d})\mc{H}_{d}(A)^{-s/d}$.
 
We next show that $A$ also satisfies the hypotheses of Proposition~\ref{prop2}
which will then imply that $\g(A)$ exists and is given by 
\beq \label{gmanA}
\g(A)=\g(U^{d})\mc{H}_{d}(A)^{-s/d}.
\eeq
Since $A$ is a $d$-rectifiable manifold, we have
    $A=\bigcup_{k=1}^{n}\phi_{k}(K_{k})$
    where $K_{k}\subset\R^{d}$ is compact
    and $\phi_{k}$ is bi-Lipschitz on $K_{k}$ with constant $L_{k}$ for $k=1,\ldots,n$.  
Let  $L:=\max\{L_{k}\mid k=1,\ldots,n\}$.  
 
 Suppose $B$ is a compact subset of  $A$. For $k=1,\ldots,n$, let 
 $B_{k}:=B\cap \phi_{k}(K_{k})$.  Then by Lemma~\ref{unionlemmalower} and Theorem~\ref{main1}
 \begin{align}
     \underg(B)&\ge \left( 
 \sum_{k=1}^{n}\underg(B_{k})^{-d/s}\right)^{-s/d} \\
 &\ge  L^{-s}\left( \sum_{k=1}^{n}\underg(\phi_{k}^{-1}(B_{k}))^{-d/s}\right)^{-s/d} 
 \nonumber \\
 &\ge L^{-s}\g(U^{d}) \left( \sum_{k=1}^{n}\mc{H}_{d}(\phi_{k}^{-1}(B_{k}))\right)^{-s/d} 
  \nonumber \\
 &\ge n^{-s/d} L^{-2s}\g(U^{d})\mc{H}_{d}(B)^{-s/d}\nonumber
\end{align}
from which it follows that $A$ satisfies the hypotheses of 
Proposition~\ref{prop2}, thereby proving (\ref{gmanA}).

Once we have the formula (\ref{gmanA}), the proof of 
Theorem~\ref{main2} may be repeated without change to show that 
(\ref{wconv}) holds for  asymptotically  optimal $s$-energy $N$-point  configurations in $A$.
   
   Finally, to  prove the 
   separation estimates (\ref{sepcond}) for 
   an optimal $N$-point   $s$-energy configuration $\lambda^{*}_{N}=\{y_{1,N}^{*},\ldots,y_{N,N}^{*}\}$
   for $A$, when $A=\phi(K)$, $K\subset\R^{d}$, $K$ compact and $\phi$ 
   bi-Lipschitz on $K$, we can imitate the argument given in 
   Section~6.1 for the proof of Theorem~\ref{main4}. For this purpose we 
   replace the definition of $U_{i}(x)$ in (\ref{Uidef}) by 
   $$
   U_{i}(x):=\sum_{j\neq i}\frac{1}{|\phi(x)-\phi(x_{j,N})|^{s}},
   $$
    where $x_{j,N}=\phi^{-1}(y_{j,N}^{*})$.  The details are left to 
    the reader.  
   \eproof
  
% In this paper, we follow Federer  \cite{Fed} and  say that a set 
% $A$ is
% is { \em $d$-rectifiable} if 
% We warn the reader that there are several 

\textbf{Acknowledgements.} \ We thank the referees,  S. Borodachov, and J. Brauchart for their careful reading of the manuscript and their helpful comments.

       \end{document}

\maketitle

\begin{appendix}
\section{Comments on the proof of Proposition   8.3}
In the proof of Proposition 8.3   the assertion was made that each $\psi_i(K_i)$  was contained in some open (relative to $A$) set $G_i$ such that (83) holds.   The case $s = d$ follows from the hypothesis that $A$ is contained in a $d$-dimensional $C^1$ manifold since in this case the sets $K_i$ may be chosen to be almost clopen relative  to  $A$. The case $s > d$ is more delicate and requires an argument similar to that used in the proof of Theorem 2.1 given in Section 7. In this addendum we provide a sketch of these arguments for the interested reader.   

In this note we let $B_d(a, r)$ denote the open ball in $R^d$ with center $a$ and radius $r$,
$\bar B_d(a, r)$   the closure of $B_d(a, r)$, and $C_d(a, r)$ the boundary of $B_d(a, r)$.

\subsection{Case 1: $s = d$}

Let $M\subset \R^{d'}$ be a $C^1$ manifold (without boundary), i.e. for each point $p\in  M$ there is some  open  set (relative to $M$)
 $V\subset M$ containing $p$,  some open ball $B_d(q,\rho)$, $\rho>0$, and a diffeomorphic mapping $\phi_p$ from $B_d(q,\rho)$ onto  $V$ (i.e. $\phi_p$ is a homeomorphism of $B_d(q,\rho)$ onto  $V$ and  $\phi_p$ is continuously differentiable on $B_d(q,\rho)$).  By composing $\phi_p$ with an appropriate affine mapping  on $\R^d$ we may assume without loss of generality  that $\phi_p(0)=p$ and that
$\phi_p'(0)^T\phi_p'(0)=I$ so that $\phi_p'(0)$ is an isometric mapping from $\R^d$ onto its range in $\R^{d'}$.  For $x,y\in B_d(0,\rho)$, let $[x,y]$ denote the directed line segment from $x$ to $y$.  Then
\begin{equation}\label{eq1}
\phi_p(y)-\phi_p(x)=\int_{[x,y]}\phi_p'(z)dz=\phi_p'(0)(y-x)+\int_{[x,y]}(\phi_p'(z(t))-\phi_p'(0))dz(t). 
\end{equation}
Let $\epsilon>0$.  From the continuity of $\phi_p'$ and (\ref{eq1}) it follows that there is some $0<\nu_p<\rho$ such that $\phi_p$ is bi-Lipschitz on $B_d(0,\nu_p)$ with constant $(1+\epsilon)$.  
Since $\phi_p(B_d(p, \nu_p))$ is a neighborhood of $p$, there is some $\delta_p>0$  such that   $\bar B_{d'}(p,\delta_p)\cap M\subset \phi_p(B_d(p, \nu_p))$.   
Furthermore, as we remark in the proof of Proposition 8.3,  for all but a countable collection of $0<\delta<\delta_p$, the set  $B_{d'}(p,\delta)\cap M$ is  $\mc{H}_d$-almost clopen.  

Now suppose $A$ is a compact subset of $M$ and consider the collection $$\mathcal{C}:=\{\bar B_{d'}(a,\delta)\cap A \mid a\in A, \delta<\delta_a ,  \mc{H}_d(C_{d'}(a,\delta)\cap A)=0 \}.$$
Applying the Besicovitch Covering Theorem we may then choose  $(\psi_i,K_i)$ in the proof of Proposition 8.3 where each $K_i$ is of the form $\phi_a^{-1}(\bar B_{d'}(a,\delta)\cap A)$ with $\psi_i=\phi_a$.
and $\psi_i(K_i)$ is almost clopen with respect to $\mc{H}_d|_A$.  We may assume without loss of generality  that each $\mc{H}_d(\psi_i(K_i))>0$ for all $i$.   

 As described in the proof of Proposition 8.3 (see the discussion following (82)) we can find for $\gamma>0$ an open and almost clopen set $G_i$ that contains $\psi_i(K_i)$ and 
is such that  $\mc{H}_d(G_i \setminus \psi_i(K_i))<\gamma.$  Since $\partial(G_i \setminus \psi_i(K_i))\subset   \partial G_i \cup \partial\psi_i(K_i)$ is a set of $\mc{H}_d$-measure 0, we have
$\mc{H}_d(G_i \setminus \psi_i(K_i))=\mc{H}_d(\overline{ G_i \setminus \psi_i(K_i)})$.  By the hypotheses of Proposition 8.3, we may choose $\gamma$ small enough so that 
\begin{equation}\label{eq2}
\g(\bar G_i \setminus \psi_i(K_i))^{-d/s}= \g(\overline{ G_i \setminus \psi_i(K_i)})^{-d/s}\le \epsilon/2^i.
\end{equation}
  By  Lemma 3.2 we have
$$
\g(\bar G_i)\ge (\g(\psi(K_i)^{-d/s}+\g(\bar G_i \setminus \psi_i(K_i))^{-d/s}) 
$$
which combined with (\ref{eq2}) gives (83).  

\subsection{Case 2: $s > d$}

 A closed set $\tilde A\subset \R^{d'}$ is said to be {\bf $d$-regular} if there are positive constants $c_0, c_1$ and $R$ such that $ c_0r^d \le \mc{H}_d(\tilde A\cap B_{d'}(a,r)) \le c_1r^d$ for all $a\in \tilde A$ and $0<r\le R$.  The assumption that $A$ is a compact subset of $B:=\bigcup_{k=1}^n \phi_k(G_k)$ where each
   $\phi_k$ is bi-Lipschitz on an open set $G_k\subset\R^d$ shows that $A\subset \tilde A\subset B$ for some $d$-regular set $\tilde A\subset \R^{d'}$.  We may assume without loss of generality that   $\underg(\psi_i(K_i))>0$, $i=1,2,\ldots$.
Applying the following lemma~\ref{RegLemma} to each $\psi_i(K_i)$ then shows the existence of some  open (relative to $A$) set $G_i\supset \psi_i(K_i)$   such that (83) holds.

%For example, if $A=\bigcup_{k=1}^n\phi_k(K_k)$ for bi-Lipschitz mappings 
%$\phi_k$ that are bi-Lipschitz on open sets $G_k\supset K_k$, then $A$ is regular. 

\begin{lemma}\label{RegLemma}
Suppose $K\subset \R^{d'}$   is a compact subset of a $d$-regular set $\tilde A\subset \R^{d'}$.  Assume further that $s> d$ and $0<\underg(K)\le \overg(K)<\infty$.   Then for each $\epsilon>0$,  there is some $\delta>0$ such that
\begin{equation} \label{GK}
\underg(G)\ge  (1 -\epsilon)\underg(K).
\end{equation}
for any  open (relative to $ \tilde A$) set $G\supset K$ with  $\mc{H}_d(\bar G\setminus K)<\delta$.
\end{lemma}

\begin{proof}
Note that the topology we use here is the $\tilde A$-relative topology.  For convenience, we set $\B(a,r):=B_{d'}(a,r)\cap \tilde A$ for $a\in \tilde A$ and $r>0$.

\newcommand{\step}[1]{\noindent {\sc Step #1.}}

\step{1} Let $0<\delta<(1/2)^{4d}$ (we will choose $\delta$  sufficiently small at the end of this proof).  Suppose $G$  is an open set  in $\tilde A$  such that
$K\subset G$ and $\mc{H}_d(\bar G\setminus K)<\delta$.  Since $\mc{H}_d|_{\tilde A}$ is a Radon measure, there is an open set $V\subset \tilde A$,
with  $ \bar G \subset V$ such that   $\mc{H}_d(V\setminus \bar G)<\delta$. Let 
$\gamma:=\dist (\bar G,\tilde A\setminus V)$ and note that $\gamma>0$ (we take $\gamma=\infty$ if $V=\tilde A$).  

\step{2} Let \{$\npts$\} denote a sequence of optimal $s$-energy configurations for $\bar G$ and let
$r_i(\npts):=\min_{x_j\in \npts\setminus \{x_i\}}|x_i-x_j|$.  For $\rho>0$, let 
$$\npts^1:=\{x_i\in\npts \mid r_i(\npts) \ge \rho N^{-1/d}\}$$
and let $\tnpts^1:=\npts\setminus \npts^1$.  Then 
\begin{equation}\label{est1}
E_s(\npts)\ge \sum_{x_i\in\tnpts^1}(\rho N^{-1/d})^{-s}= |\tnpts^1|\rho^{-s} N^{s/d}\qquad  (N\ge 2).
\end{equation}
Let $k>\overg(K)$.  Then $k > \overg(G)$, and so there is some $N_1$ such that 
$E_s(\npts)\le kN^{1+s/d}$ for $N\ge N_1$ and we deduce from (\ref{est1}) that
\begin{equation}
\frac{|\tnpts^1|}{N}\le k \rho^s\qquad (N\ge N_1).
\end{equation}

\step{3}
Let $0<\nu<1/2$ and set 
$$
\npts^2:=\{x_i\in \npts^1\mid \B(x_i, \nu\rho N^{-1/d}) \cap K \neq\emptyset\}
$$
and let $\tnpts^2:=\npts^1\setminus \npts^2$. 
Let $N_2\ge N_1$ be large enough so that  $\rho N^{-1/d} \le \min (\gamma, R)$. 
For $N>N_2$ and points $x_i$, $x_j\in\npts^1$ with $x_i\neq x_j$  
we  have $$\B(x_i,\rho N^{-1/d})\subset V \quad \text{and} \quad \B(x_i,\nu\rho N^{-1/d})\cap \B(x_j,\nu\rho N^{-1/d})=\emptyset. $$
Notice that the balls  $\B(x_i,\nu \rho N^{-1/d})$, $x_i\in \tnpts^2 $, are pairwise disjoint and contained in $V\setminus K$ and hence we have from the $d$-regularity assumption, 
\begin{align*}
|\tnpts^2| c_0\nu^d\rho^d N^{-1}&\le \sum_{x_i\in \tnpts^2}\mc{H}_d(\B(x_i,\nu\rho N^{-1/d}))= \mc{H}_d\left(\bigcup_{x_i\in \tnpts^2}\B(x_i,\nu\rho N^{-1/d})\right) \\ &\le \mc{H}_d(V\setminus K)\le 2\delta, 
\end{align*}
and so 
\begin{equation}
\frac{|\tnpts^2|}{N} \le \frac{2\delta}{c_0 (\nu \rho)^d}.
\end{equation}
Thus
\begin{equation}
\frac{|\npts^2|}{N}=\frac{N-|\tnpts^1|-|\tnpts^2|}{N}\ge 1-k\rho^s-\frac{2\delta}{c_0(\nu\rho)^d}. 
\end{equation}

\step{4}
For $x_i\in\npts^2$, there must be some $y_i\in K$ such that $|x_i-y_i|<\nu\rho N^{-1/d}$.  Let $\npts^K:=\{y_i\mid x_i\in \npts^2\}$.
For $N>N_2$, and  $x_i, x_j\in\npts^2$, $x_i\neq x_j$, we   have $$|x_i-y_i|\le \nu\rho N^{-1/d}\le \nu |x_i-x_j|,$$ and so
\begin{equation}\label{ydist}
|y_i-y_j|=|y_i-x_i+x_i-x_j+x_j-y_j|\ge |x_i-x_j|-|y_i-x_i|-|y_j-x_j|\ge (1-2\nu)|x_i-x_j|,
\end{equation}
 which shows that $|\npts^K|=|\npts^2|$.

\step{5}
      From (\ref{ydist}),  we have $|x_i-x_j|\le (1-2\nu)^{-1}|y_i-y_j|$ for $x_i, x_j\in\npts^2$ and so, for $N>N_2$, we have
$$
\frac{E_s(\npts)}{N^{1+s/d}} \ge (1-2\nu)^s \frac{E_s(\npts^K)}{N^{1+s/d}} \ge (1-2\nu)^s \left(\frac{|\npts^2|}{N}\right)^{1+s/d}\frac{E_s(\npts^K)}{|\npts^2|^{1+s/d}}. 
$$
Taking the limit inferior as $N\to \infty$ then gives
$$
\underg(G)=\liminf_{N\to \infty} \frac{E_s(\npts)}{N^{1+s/d}}\ge (1-2\nu)^s(1-k\rho^s-\frac{2\delta}{c_0(\nu\rho)^d})_+^{1+s/d}\underg(K)
$$
where $x_+=\max (x,0)$.  
Choosing $\rho=\nu=\delta^{1/(4d)}=$ gives 
 $$
\underg(G)\ge (1-2\delta^{1/(4d)})^s(1-k\delta^{s/(4d)}- 2\delta^{1/2}/c_0)_+^{1+s/d}\underg(K).
$$

Finally, if $\epsilon>0$, then we can choose $\delta>0$ sufficiently small so that  (\ref{GK}) holds. 
\end{proof}

\end{appendix}